\documentclass[12pt,showpacs,nofootinbib]{revtex4-1}
\usepackage{multirow}
\usepackage{amssymb}
\usepackage{tikz}
\usepackage{amssymb}
\usepackage{amsmath,graphicx}
\usepackage[compat=1.1.0]{tikz-feynman}

\usepackage{array}
\usepackage{dcolumn}
\usepackage{braket}
\usepackage{bm}
\usepackage{enumerate}
\usepackage{slashed}
\usepackage{epstopdf}
\usepackage{amsmath}
\usepackage{dcolumn,caption,booktabs}
\usepackage{slashed}
\usepackage{amsfonts}
\usepackage{adjustbox}
\usepackage{tcolorbox}
\usepackage{float}
\usepackage{collectbox}
\usepackage{listings}
\usepackage[unicode]{hyperref}
\usepackage[mathscr]{eucal}
\usepackage{graphics}
\usepackage{subcaption}
\newcommand{\be}{\begin{equation}}
	\newcommand{\ee}{\end{equation}}
\newcommand{\bea}{\begin{eqnarray}}
	\newcommand{\eea}{\end{eqnarray}}
\newcommand{\ben}{\begin{enumerate}}
	\newcommand{\een}{\end{enumerate}}
\newcommand{\bde}{\begin{widetext}}
	\newcommand{\ede}{\end{widetext}}
\newcommand{\nn}{\nonumber}
\newcommand{\crn}{\nonumber \\}

\newcommand{\al}{\alpha}
\newcommand{\la}{\lambda}

\newcommand{\ga}{\gamma}
\newcommand{\va}{\varphi}

\newcommand{\pa}{\partial}
\newcommand{\+}{\dagger}
\newcommand{\fr}{\frac}

\newcommand{\bc}{\begin{center}}
	\newcommand{\ec}{\end{center}}
\newcommand{\Ga}{\Gamma}
\newcommand{\de}{\delta}
\newcommand{\De}{\Delta}

\newcommand{\varep}{\varepsilon}
\newcommand{\ka}{\kappa}
\newcommand{\La}{\Lambda}

\begin{document}
	\newcommand{\AdrHEPC}{$^a$Department of Theoretical Physics, University of Science, Ho Chi Minh City 70000, Vietnam\\ $^b$Vietnam National University, Ho Chi Minh City 70000, Vietnam}
	
	\title{Dual electroweak phase transition in the two-Higgs-doublet model with the $S_3$ discrete symmetry}
	\author{Vo Quoc Phong$^{a,b}$}
	\email{vqphong@hcmus.edu.vn}
	\affiliation{\AdrHEPC}
	\author{Nguyen Minh Anh$^{a,b}$}
	\email{minhanhkhtn@gmail.com}
	\affiliation{\AdrHEPC}
	\author{Hoang Ngoc Long$^{c,d}$
	}
	\email{hoangngoclong@vlu.edu.vn (corresponding author)}
	\affiliation{$^c$  Subatomic Physics Research Group,
Science and Technology Advanced Institute,
Van Lang University, Ho Chi Minh City 70000, Vietnam
	\\
	$^d$ Faculty of Applied Technology, School of  Technology,  Van Lang University, Ho Chi Minh City 70000, Vietnam}
		
	\begin{abstract}
In this work, dual electroweak phase transition (EWPT) consisting of two phases, is carefully studied  in the two-Higgs-doublet model with the $S_3$ discrete symmetry. The role of $S_3$ here is to further separate the stages of the electroweak phase transition, compared to that of the original two-Higgs-doublet model (2HDM). The strength of the electroweak phase transition $(S)$ in  the model under consideration is large enough for the first-order EWPT, specifically $1 < S < 2.8$. The ratio between the two vacuum expectation values (VEVs), $\tan\beta = v_2/v_1$, is proven to have no effects on the strength of the phase transition. This ratio only affects the mass domain that causes the first-order phase transition. Furthermore, in this paper we will show clearly that when studying the EWPT in models of more than one scalar field that generates masses, one needs to analyze the problem of phase transition under multiple stages. In other words, the effect of the first stage of symmetry breaking to the second one, is to simplify by suggestion that vacuum expectation value of the Higgs boson responsible for the initial stage is proportional to that of the field for the next stage.
\end{abstract}
	\maketitle
	\tableofcontents
\section{INTRODUCTION}\label{secInt}
	
The \textit{Standard Model} (SM), an outstanding achievement of physics in particular, and a memorable milestone for the scientific community in general is a systematic theory of elementary particles and their interactions. The model predicted the results of many experiments, the existence of the Higgs particle; together with the Higgs mechanism, it shows us the nature of subatomic particles. However, the model still has some shortcomings such as not being able to unify gravity, describe dark matter or small neutrino mass, etc.
	
One of the significant phenomena in cosmology that cannot be explained by the SM is \textit{baryon asymmetry}, also known as matter-antimatter asymmetry. This problem is explains why there is an imbalance between matter and antimatter in the universe. For a strong \textit{first-order electroweak phase transition}, the third of Sakharov's three conditions \cite{sakharov}, plays an important role in explaining this asymmetry. It indeed important because this condition not only explains the thermal imbalance but also provides a link between the violation of B and \textit{CP} and the other two conditions of Sakharov.
	
The thermal imbalance is expressed through a first-order electroweak phase transition (EWPT) which should be considered first. The SM does not have enough triggers for a first-order phase transition \cite{mkn,SME,SMEb,SMEc,SMEd,michela}. Therefore, in the beyond  SM, this problem  must be considered (see, for example,  Refs.~\cite{plv,2b,2c,BSM,BSMb,majorana,majoranab,thdm,thdmb,ESMCO,elptdm,elptdma,elptdmb,elptdmc,elptdmd,phonglongvan,phonglongvanb,phonglongvan2,SMS,dssm,munusm,lr,singlet,singletb,singletc,singletd,mssm1,mssm1b,mssm1c,twostep,twostepb,twostepc,1101.4665,1101.4665b,1101.4665c,jjgb,jjgc,jjgd,Ahriche1,Ahriche2,Ahriche2b,Ahriche3,Ahriche3b,Ahriche3c,Ahriche3d,Ahriche3e,Fuyuto,Fuyutob,Fuyutoc,span,chr,cde,kusenko}). The different scenarios that can be enumerated in these references are as follows: doing high-temperature effective potential, analysis of the trigger roles of new particles, the decoupling conditions, bubble nucleations, and sphalerons.
	
The triggers for the EWPT can be new particles (beyond SM) or parameter corrections in SM \cite{plv,2b,2c,BSM,BSMb,majorana,majoranab,thdm,thdmb,ESMCO,elptdm,elptdma,elptdmb,elptdmc,elptdmd,phonglongvan,phonglongvanb,phonglongvan2,SMS,dssm,munusm,lr,singlet,singletb,singletc,singletd,mssm1,mssm1b,mssm1c,twostep,twostepb,twostepc,1101.4665,1101.4665b,1101.4665c,jjgb,jjgc,jjgd,Ahriche2,Ahriche2b,Ahriche3,Ahriche3b,Ahriche3c,Ahriche3d,Ahriche3e,Fuyutob,Fuyutoc,span,chr,cde,kusenko}. In the SM, it is contrary to the experiment that the mass of Higgs boson must be less than $125$ GeV, for a strong first-order EWPT \cite{mkn,SME,SMEb,SMEc,SMEd,michela}. In a model, if the new particles are the cause of the violent EWPT, then that model can have more than one Higgs field \cite{plv,2b,2c,majorana,majoranab,thdm,thdmb,ESMCO,elptdm,elptdma,elptdmb,elptdmc,elptdmd,phonglongvan,phonglongvanb,phonglongvan2,epjc,zb,singlet,singletb,singletc,singletd,mssm1,mssm1b,mssm1c,twostep,twostepb,twostepc,chiang3}.	Another interesting point but consistent with the physical nature, the strength of EWPT is gauge-independent \cite{zb, 1101.4665, 1101.4665b,1101.4665c,Arefe}. The self-energy term or daisy loops cause a problem for effective potentials at high temperatures. However, it is not the main trigger for EWPT and it reduces the strength of EWPT \cite{r23}.
	
By making sure that the \textit{C} and \textit{CP} violations exist, the third condition of baryogenesis is given as $\Ga_{sph}\sim \mathcal{A}(T) \textrm{exp}\left(\fr{- E_{sph}}{T}\right) \ll H_{rad}$ \cite{spha-huble,decoupling, decouplingb, decouplingc} in the context of topological transitions, where $\Ga_{sph}$ and $E_{sph}$ are the sphaleron rate and energy, respectively, $H_{rad}$ is the Hubble expansion rate in the radiation-dominated period, and $\mathcal{A}(T)\approx T^4$. This is referred to as the sphaleron decoupling situation. This condition is frequently written as $S=v_c/T_c>1$ in the SM using the approximation $E_{sph}(T)\approx [v(T)/v] E_{sph}(T=0)$ \cite{47, Ahriche1, Ahriche2,Ahriche2b}. However, this approximation should be used with caution in models that go beyond the SM.
	
Through the above brief summaries, ones have  another aspect of EWPT survey. Currently, there are two scenarios as follows: the first one is that the EWPT process has only one stage; the second is that this process has two or three stages. Some of the theoretical models that study one stage of EWPT are the SM, Zee-Babu \cite{zb}, SMEFT \cite{pkll}, SM with corrections to the Yukawa interaction for quarks \cite{yukawa1,yukawa2}. Models that study more than one stage of EWPT are the ones that consist of more than one Higgs field. Some of them as follows: 3-3-1 models \cite{plv,phonglongvan2,ptl}, 2-2-1 model \cite{pa}.	
	
However, there is one quite special model, the 2HDM has two vacuum expectation values (VEVs) having values/ranges  in the electroweak scale. In this paper, the ways of studying the EWPT with only one stage of this model will be reconsidered. Then the strengths and weaknesses of it will be analyzed. At the same time, there are many interesting versions of the model that go beyond itself, in which there is a model of two Higgs doublets with symmetry of $S_3$. This symmetry can account for quark mixing \cite{bpal1,bpal2,bpal3}. The $S_3$ symmetry has been proposed as the basic flavor symmetry in various frameworks. This discrete symmetry in the lepton sector is to produce a $\mu-\tau$ symmetry \cite{28t,29t,30t,31t} or a tribimaximal neutrino mixing matrix \cite{32t,33t}. In the quark sector, this symmetry can produce Fritzsch and Fritzsch mass textures \cite{34t}. In addition, the nearest neighbor interactions (NNI) mass texture is hidden in a $S_3$ flavor symmetry \cite{35t}. Therefore, the model with $S_3$ is interested, since $S_3$ would simplify the Higgs sector, which is very important in studying EWPT. Furthermore, since there are only two VEVs, our work of studying the multi-stages of the 2HDM-$S_3$ would be relatively easier compared to that of the SM.	
	
The two-Higgs-doublet model with $S_3$ symmetry (2HDM-$S_3$) \cite{1601}, one of the extended versions the 2HDM, that has the potential to "possess" a strongly first-order electroweak phase transition, because of the following factors: the model has the heavy Higgs boson, as well as the charged Higgs boson; at the same time, it obeys the smallest non-Abelian discrete symmetry group.	
	
More specifically, we will consider whether the first-order phase transition in the 2HDM-$S_3$ is strong or not. And when it is strong, the range of values of the phase transition strength and the mass of the new particles and related parameters will be investigated.	
	
The paper has the following structure. Except for the Introduction (Sec.~\ref{secInt}) and the Conclusion and Outlooks (Sec.~\ref{vi}), Appendix~\ref{multi} and Sec.~\ref{iii} give a quick review of the effective potentials, as well as some comments and remarks on the electroweak phase transition in the 2HDM. In Sec.~\ref{iv} and Sec.~\ref{v}, the electroweak phase transition in the 2HDM-$S_3$ is studied. More specifically, the effective potential of the dual electroweak phase transition will be studied, the strength of first-order phase transition, the mass domain for the first-order phase transition, all of which will be given with a parameter $a$ that will be introduced to replace the popular parameter $\tan\beta=v_2/v_1$.

\section{Review on the Higgs potential and comments on EWPT in the 2HDM}\label{iii}
	
\subsection{The Higgs potential in the 2HDM}\label{iiia}
	
 The fermion  and scalar spectrum with their assignments under the $SU(2)_L\times U(1)_Y$ gauge group are given by ~\cite{davidson,Herrero-Garcia:2017xdu}:
\begin{align}
L_{a}&=	\begin{pmatrix}
		\nu_{aL} \\ e_{aL} 
	\end{pmatrix}
	\sim (2, -1)\, , \;\quad Q_{aL}= \begin{pmatrix}
		u_{aL}\\ d_{aL} 
	\end{pmatrix} 
	\sim \left( 2, \dfrac{1}{3}\right)\,,
	\crn 	e_{aR} &\sim (1,-2), \quad   u_{aR} \sim \left(1,\dfrac{4}{3}\right)\,, \quad  d_{aR} \sim  \left(1, -\dfrac{2}{3}\right),  \; a=1,2,3 \,, 
\crn & \Phi_i=	\begin{pmatrix}
	\phi^+_i\\\phi^0_i 
\end{pmatrix}
\sim (2, 1)\,, \;  \langle  \Phi_i \rangle =	\frac{1}{\sqrt{2}}\begin{pmatrix}
0\\
v_i
\end{pmatrix}
\sim (2, 1), i=1,2\,.
\end{align}
	
For details of the quark sector of different types of the 2HDM, the reader is referred to Refs. \cite{davidson, Herrero-Garcia:2017xdu}. 
In the 2HDM, the charge operator is defined as
\be Q = T_3 + \fr Y 2\,.
\label{l}
\ee

The most general form of the effective potential in the 2HDM contains $14$ parameters, and there may exist the \textit{CP} conservation (charge and mirror symmetry), \textit{CP} violation, and charge violation. When expressing the form of the potential, we must be careful in defining the quantities and distinguishing the parameters, since when applying the group rotations, the physics can be changed. However, in the studies of the phenomenology of the  2HDM, assumptions are often made to simplify the calculations. For example, \textit{CP} is often assumed to be conserved in the Higgs fields (only then can we distinguish the scalar field from the scalar pseudo-field). Now, the \textit{CP} discrete symmetry will eliminate all the fourth power terms that contain the odd number of one of two Higgs fields from the potential (for instance, $\Phi_1^\dagger \Phi_1 \Phi_1^\+ \Phi_2$). We can also assume that all of the parameters corresponding to the fourth power terms are of real number, including the term added to break the symmetries.	
	
With the above assumptions, putting in the symmetry $Z_2$ ($\Phi_1 \rightarrow \Phi_1$, $\Phi_2 \rightarrow - \Phi_2$) in order not to have FCNC at the tree level. After this, the most general form for the scalar potential for the two doublets $\Phi_1$ and  $\Phi_2$ with supercharge $+1$ would have the following form \cite{tdhm}:	
	
	\begin{align}
		\begin{split}
			V = &\, \, \mu_{11}^2 \Phi_1^2+ \mu_{22}^2 \Phi_2^2 - \mu_{12}^2 \left( \Phi_1. \Phi_2 + \Phi_2. \Phi_1 \right) + \dfrac{\la_1}{2}   \Phi_1^4 + \dfrac{\la_2}{2}  \Phi_2^4 \\
			& +\la_3 \Phi_1^2 \Phi_2^2 + \la_4 (\Phi_1. \Phi_2)( \Phi_2. \Phi_1) + \dfrac{\la_5}{2}   \left(  \Phi_1^4 +  \Phi_2^4 \right),\label{HiggsPotential}
		\end{split}
	\end{align}
where we have denoted $\Phi_i^2 = \Phi_i^\dag \Phi_i,\,  = 1,2 $ and $\Phi_i . \Phi_j =  \Phi_i^\dag \Phi_j , \, i\neq j$.	All the parameters  $\mu_{11}$, $\mu_{22}$, $\la_i$ $(i=1,\dots,5)$ are all real and the term containing $\mu_{12}$ “softly” breaks the symmetry $Z_2$. Following this, there are two complex scalar doublets $SU(2)$ containing eight fields,
	\begin{align}
		\Phi_a =
		\begin{pmatrix}
			\De_a^+\\
			\fr 1{ \sqrt{2}}(v_a + \rho_a + i \eta_a)
		\end{pmatrix}
		, \qquad a=1,2.
	\end{align}
	
Three of them are Goldstone bosons  eaten by the massive  gauge bosons $W^\pm$ and $Z^0$ to generate their masses. The other five are physical scalar Higgs fields, including one scalar carrying charge, two neutral scalars, and one pseudo scalar.
	
Averaging over the whole space, VEVs read
	\begin{align}
		\label{VEV}
		\langle \Phi_1 \rangle_0 =\fr 1{ \sqrt{2}}
		\begin{pmatrix}
			0\\
			v_1
		\end{pmatrix}
		,
		\langle \Phi_2 \rangle_0 = \fr 1{ \sqrt{2}}
		\begin{pmatrix}
			0\\
			v_2
		\end{pmatrix}
		,
	\end{align}
One of the  most  important parameters of the model
is as follows:
	\begin{align}
			\tan \beta =\fr{s_\beta}{c_\beta}= \dfrac{v_2}{v_1}\,,
	\end{align}
where we have  used the notations $ s_\beta\equiv \sin\beta, \, c_\beta \equiv\cos\beta$. Here $\beta$ is the rotational angle when normalizing the squared mass matrix of the charged and pseudoscalars.
	
Two general neutral even \textit{CP} states $\rho_1$ and $\rho_2$ are not physical mass states. The mass matrix corresponding to them can be diagonalized by a rotation of a mixed angle of $\rho_1$ and $\rho_2$.
	
Having two minima as above inserted into the Lagrangian which contains the mass terms corresponding to the scalar fields. First, the term that corresponds to the charged scalar field is
	
	\begin{align}
		\mathcal{L}_{\phi^\pm , mass} = \left[ \mu_{12}^2 - (\la_4 + \la_5) v_1 v_2 \right]
		\begin{pmatrix}
			\De_1^- & \De_2^-
		\end{pmatrix}
		\begin{pmatrix}
			\tan \beta &-1\\
			-1 &\coth \beta 
		\end{pmatrix}
		\begin{pmatrix}
			\De_1^+ \\
			\De_2^+
		\end{pmatrix}.
	\end{align}

After diagonalizing the above mass matrix, the squared mass of the charged Higgs particle,	
	\begin{align}
		m_{H^\pm}^2 = \left( \dfrac{\mu_{12}^2}{v_1 v_2} - \la_4 - \la_5  \right) (v_1^2 + v_2^2) = \left( \dfrac{\mu_{12}^2}{v_1 v_2} - \la_4 - \la_5  \right) v^2.\label{hatHH}
	\end{align}
Next, the mass term that corresponds to the pseudoscalar field,
	\begin{align}
		\mathcal{L}_{\eta, mass} = \left( \dfrac{\mu_{12}^2}{v_1 v_2} - \la_5  \right)
		\begin{pmatrix}
			\eta_1 & \eta_2
		\end{pmatrix}
		\begin{pmatrix}
			v_2^2 & -v_1 v_2\\
			-v_1 v_2 & v_1^2
		\end{pmatrix}
		\begin{pmatrix}
			\eta_1 \\
			\eta_2
		\end{pmatrix}.
	\end{align}
The physical squared mass of the pseudoscalar after normalization,		
	\begin{align}
		m_A^2 = \left( \dfrac{\mu_{12}^2}{v_1 v_2} - 2\la_5 \right) (v_1^2 + v_2^2).\label{hatA}
	\end{align}
Finally, the mass term that corresponds to the remaining two scalar fields is
	\begin{align}
		\mathcal{L}_{\rho, mass} = -
		\begin{pmatrix}
			\rho_1 & \rho_2
		\end{pmatrix}
		\begin{pmatrix}
			\mu_{12}^2 \fr{v_2}{v_1} + \la_1 v_1^2 & -\mu_{12}^2 + \la_{345} v_1 v_2\\
			-\mu_{12}^2 + \la_{345} v_1 v_2 & \mu_{12}^2 \fr{v_1}{v_2} + \la_2 v_2^2
		\end{pmatrix}
		\begin{pmatrix}
			\rho_1 \\
			\rho_2
		\end{pmatrix}.
\label{27}	\end{align}
	
With $\la_{345} = \la_3  + \la_4 + \la_5 $, after normalizing the matrix in Eq.\eqref{27}, the squared masses of the light $(h)$ and the heavy $(H)$ Higgs particles are, respectively:
	
	\begin{align}
		\begin{split}
			& m_h^2 = \dfrac{1}{2} \left[ (\la_1 v_1^2 + \la_2 v_2^2) + \mu_{12}^2 \dfrac{v^2}{v_1 v_2} \right] \\
			& \qquad - \sqrt{\left[ \dfrac{\la_1 v_1^2 - \la_2 v_2^2}{2} - \dfrac{\mu_{12}^2}{2 v_1 v_2} (v_1^2 - v_2^2)\right]^2 + (\la_{345} v_1 v_2 - \mu_{12}^2)^2},\\
			& m_H^2 = \dfrac{1}{2} \left[ (\la_1 v_1^2 + \la_2 v_2^2) + \mu_{12}^2 \dfrac{v^2}{v_1 v_2} \right] \\
			&\qquad + \sqrt{\left[ \dfrac{\la_1 v_1^2 - \la_2 v_2^2}{2} - \dfrac{\mu_{12}^2}{2 v_1 v_2} (v_1^2 - v_2^2)\right]^2 + (\la_{345} v_1 v_2 - \mu_{12}^2)^2},\label{hatH}
		\end{split}
	\end{align}	
	
From the Eqs.~\eqref{hatHH}, \eqref{hatA}, \eqref{hatH}, the squared masses of the Higgs particles always contain in them troublesome mixing terms of VEVs. A quantity is introduced as follows:
	
	\be
	M^2_{Higgs}(v_1,v_2)=m^2_H H^2+ m^2_h h^2+m^2_A A^2+m^2_{H^\pm}(H^\pm)^+H^\pm.\label{11}
	\ee
	
\subsection{Higgs potential at the tree level}\label{iiib}
	
	From the Higgs potential given in  Eq.~\eqref{HiggsPotential}, $V_{0}$  in a form that is dependent on the VEVs as follows:
	\be\label{V0}
		V_{0}(v_1,v_2)= \fr{\mu_{11}^2}{2} v_1^2+ \fr{\mu_{22}^2}{2} v^2_2 - \fr{\mu_{12}^2}{2} v_1v_2 + \left( \dfrac{\la_1}{8}+\dfrac{\la_5}{8} \right) v_1^4 + \left(\dfrac{\la_2}{8}+\dfrac{\la_5}{8}\right)  v_2^4 +\left(\fr{\la_3}{4} + \fr{\la_4}{4}\right)v_2^2v_1^2.
	\ee	
	$V_{0}(v_1,v_2)$ has a quartic form like in the SM. On the other hand, by developing the Higgs potential Eq.~(\ref{HiggsPotential}), two minimum equations which permit us to transform the mixing between $v_1$ and $v_2$,
	\bea
	\fr{\pa V}{\pa v_1}&=&0, \mu_{11}^2v_1-\mu_{12}^2v_2+\left(\la_1+\la_5\right)\fr{v^3_1}{2}+\left(\la_3 + \la_4\right)\fr{v_2^2v_1}{2}=0.\crn
	\fr{\pa V}{\pa v_2}&=&0, \mu_{22}^2v_2-\mu_{12}^2v_1+\left(\la_2+\la_5\right)\fr{v^3_2}{2}+\left(\la_3 + \la_4\right)\fr{v_2v_1^2}{2}=0.
	\label{eqn17Feb}
	\eea
	From Eq.~(\ref{eqn17Feb}), a relationship between VEVs, such as
	\be
		\mu_{22}^2v^2_2+\left(\la_2+\la_5\right)\fr{v^4_2}{2}=\mu_{12}^2v_1v_2-\left(\la_3+\la_4\right)\fr{v_2^2v_1^2}{2}.\label{c}
	\ee
	
	Substituting Eqs.~(\ref{c}) into  Eq.~(\ref{V0})  yields
	\be\label{V0s}
		V_{0}(v_1,v_2)= \fr{\mu_{11}^2}{2} v_1^2+ \left(\dfrac{\la_1}{8}+\dfrac{\la_5}{8} \right) v_1^4 - \left(\dfrac{\la_2}{8}+\dfrac{\la_5}{8}\right)  v_2^4=V_0(v_1)+V_0(v_2),
	\ee	
	where  $V_0(v_1)=\fr{\mu_{11}^2}{2} v_1^2+ \left(\dfrac{\la_1}{8}+\dfrac{\la_5}{8} \right) v_1^4$ and $V_0(v_2)=\left(-\dfrac{\la_2}{8}-\dfrac{\la_5}{8}\right)v_2^4$ are in the quartic form. In addition, there are alternative ways to arrive Eq.~\eqref{V0s} which has other forms but $V_0(v_1)$ and $V_0(v_2)$ are still in the quartic form.	
	
	If the potential at the tree level had the quartic form of each vacuum expectation value, that is there are no mixing terms in it, the job of calculating the effective potential for each VEV would be much easier. This will be made clear in the following sections.

	\subsection{The masses of gauge bosons}\label{iiic}
	
	In order to find the gauge boson masses, we starting from the kinetic term of the Higgs fields. In the 2HDM-$S_3$, there are two components in the kinetic term for the two Higgs doublets,
	
	\be
	\mathcal{L}^{GB}_{mass}=\left( \mathcal{D}_{\mu }\langle \Phi_1\rangle \right) ^{\+ }\left( \mathcal{D}^{\mu }\langle \Phi_1\rangle \right) +\left( \mathcal{D}_{\mu }\langle \Phi_2\rangle \right)
	^{\+ }\left( \mathcal{D}^{\mu }\langle \Phi_2\rangle \right) \equiv  A+B,
	\ee
	in which, the covariant derivatives act on $\Phi_1$ and $\Phi_2$.
  as follows
\be  \mathcal{D}^{\mu }  = \pa^\mu - i g W^\mu_i T_i - \fr i 2 g^\prime B^\mu Y
\,.
\ee
	Note that the gauge fields ($W^\mu_i, B^\mu$) inside the covariant derivatives of $A$ and $B$  are the same. So after diagonalizing, the gauge fields in $A$ and $B$ are the same, and gauge bosons $\ga $, $Z$, $W^{\pm}$ are inferred.
	
	From the term $A$, one obtains the mass components of the physical gauge bosons only depends on $v_1$,
	\be\label{ma}
		\begin{split}
			M^A_{bosons}=m^2_{W^{\pm}}(v_1)W^{+}_{\mu}W^{-\mu}+m^2_{Z}(v_1)Z_{\mu}Z^{\mu}.
		\end{split}
	\ee
	
	From the term $B$, one obtains the mass components of the physical gauge bosons that only depend on $v_2$,
	\be\label{mb}
		\begin{split}
			M^B_{bosons}=m^2_{W^{\pm}}(v_2)W^{+}_{\mu}W^{-\mu}+m^2_{Z}(v_2)Z_{\mu}Z^{\mu}.
		\end{split}
	\ee
	
	Therefore the bosons masses can be split into two parts,
	\be\label{BosonMassSplit}
		m^2_{gauge-boson}(v_1, v_2)= m^2_{boson}(v_1)+m^2_{boson}(v_2).
	\ee
	
	Similar arguments for the problem can be found in Ref.~\cite{phonglongvan2}.
	
	\subsection{Remarks on EWPT structure in the 2HDM}
	
	The Higgs and gauge boson sectors from the full Higgs Lagrangian,
	\be \label{HiggsLagrangian}
		\mathcal{L}=\mathcal{L}^{GB}_{mass} - V(\Phi_1,\Phi_2),
	\ee
	where $V(\Phi_1,\Phi_2)$ is given by Eq.~\eqref{HiggsPotential}.
	
	Expanding the Higgs fields $\Phi_1$ and $\Phi_2$ around their VEVs which are $v_1$, $v_2$, yields 
	\bea \label{HiggsLagrangian-1}
	\begin{split}
		\mathcal{L}
		=&\fr{1}{2}\pa^{\mu}v_1\pa_{\mu}v_1+
		\fr{1}{2}\pa^{\mu}v_2\pa_{\mu}v_2 - V_0(v_1,v_2)
		+M^A_{gauge-bosons}+M^B_{gauge-bosons}\\
		&+M^2_{Higgs}(v_1,v_2)+[m_{top-quark}(v_1)+m_{top-quark}(v_2)]t\bar{t}.
	\end{split}
	\eea
	Therefore, from the Lagrangian in Eq.~\eqref{HiggsLagrangian-1}, two motion equations according to $v_1$ and $v_2$ are calculated,
	\bea
	\pa^{\mu}v_1\pa_{\mu}v_1 - \fr{\pa V_0(v_1)}{\pa
		v_1}+\sum \fr{\pa m^2_{bosons}(v_1)}{\pa v_1}W^{\mu}
	W_{\mu}+\fr{\pa m_{top}(v_1)}{\pa v_1}t\bar{t}+\fr{\pa M^2_{Higgs}(v_1,v_2)}{\pa v_1} &= &0,\label{mchi1a}\\
	\pa^{\mu}v_2\pa_{\mu}v_2 -  \fr{\pa V_0(v_2)}{\pa v_2}
	+\sum \fr{\pa m^2_{bosons}(v_2)}{\pa v_2}W^{\mu}W_{\mu}+
	\fr{\pa m_{top}(v_2)}{\pa v_2}t\bar{t}+\fr{\pa M^2_{Higgs}(v_1,v_2)}{\pa v_2}&=& 0, \label{mrho1}
	\eea
	where $W$ runs over all gauge fields. 	
	
	If $M^2_{Higgs}(v_1,v_2)$ is like Eq.~(\ref{11}), in which there are no mixing terms of VEVs, this term can be separated into two terms such that each of the new terms only depends on one VEV. However, the fifth  Higgs particles’ masses in the 2HDM all have mixing terms.	
	
	Next, there are an important observation, that when the universe was cooling down to the value of $v_2$ after the big bang, the field $\Phi_2$ broke the electroweak symmetry, and after that when the universe continued to cool down to the value of $v_1$, the field  $\Phi_1$ continued to break the electroweak symmetry once again. The process of this electroweak symmetry breaking must be sequential.  Hence, we cannot combine the two stages to study.
	
	Therefore the rules for generating the masses of the particles through two symmetry breakings as follows:
	
	\textit{Remark 1}: At stage 1, $\Phi_2$ breaks the symmetry or $v_2\ne 0$, but now $\Phi_1$ has not yet broken the symmetry so $v_1$ is still equal to 0. Hence, all the Higgs particles’ masses only contain $v_2$.
	
	\textit{Remark 2}: When the breaking symmetry occurs at $\Phi_1$, the interactions between $\Phi_2$ and $\Phi_1$ are turned on and $v_1$ would not be 0. In this stage, the further generated masses can not only depend on $v_2$, but also on $v_1$ and the mixing of $v_2$ and $v_1$.
	
	\textit{Remark 3}: With the above two remarks, through the mixing terms in the masses of the Higgs particles, the effects of the first stage has on the second stage. But they also make it difficult for investigating the phase transition at the later stage.
	
	\textit{Remark 4}: In order to view the two  phase transition stages with the separated effective potentials, we can apply the following approximation to the mixing terms: $v_1.v_2\sim \kappa.v_1^2$. Since at this time, $v_2$ can still change as the temperature decreases, it can consider the change of $v_2$ is now equal to $\kappa v_1$. This remark is actually a mathematical treatment like the approximation $v^2=v_1^2+v^2_2$. But when it is combined with the above third remarks, they make physical sense in the analysis of EWPT.
	
	Therefore, from four remarks, all the Higgs particles’ masses can be split into two different components,
	\be
	M^2_{Higgs}(v_1,v_2)=m^2_{Higgs}(v_1)+m^2_{Higgs}(v_2).
	\ee

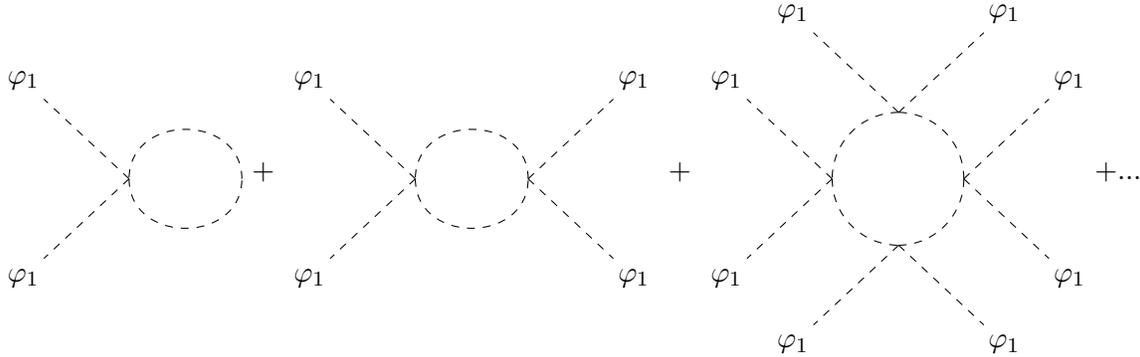
\begin{figure}[h!]
		\begin{tikzpicture}[baseline=(current  bounding  box.center)]
			\begin{feynman}
				\vertex (x);
				\vertex[right=1.5cm of x] (y);
				\vertex[above left=of x] (a){\(\va_1\)};
				\vertex[below left=of x] (b){\(\va_1\)};
				\vertex[above right= 0.5cm and 0.75cm of x] (p);
				
				\diagram*{
					(y) --[scalar, half right] (x),
					(x) --[scalar, half right] (y),
					(x) --[scalar] (a),
					(x) --[scalar] (b),
						};
			\end{feynman}
		\end{tikzpicture}
		+
		\begin{tikzpicture}[baseline=(current  bounding  box.center)]
			\begin{feynman}
				\vertex (x);
				\vertex[right=1.5cm of x] (y);
				\vertex[above left=of x] (a){\(\va_1\)};
				\vertex[below left=of x] (b){\(\va_1\)};
				\vertex[above right=of y] (c){\(\va_1\)};
				\vertex[below right=of y] (d){\(\va_1\)};
				
				\diagram*{
					(y) --[scalar,  half right] (x),
					(x) --[scalar, half right] (y),
					(x) --[scalar] (a),
					(x) --[scalar] (b),
					(y) --[scalar] (c),
					(y) --[scalar] (d),
				};
			\end{feynman}
		\end{tikzpicture}
		+
		\begin{tikzpicture}[baseline=(current  bounding  box.center)]
			\begin{feynman}
				\vertex (x);
				\vertex[right=1.75cm of x] (y);
				\vertex[above right=1.25cm of x] (z);
				\vertex[below right=1.25cm of x] (k);
				\vertex[above left=of x] (a){\(\va_1\)};
				\vertex[below left=of x] (b){\(\va_1\)};
				\vertex[above right=of y] (c){\(\va_1\)};
				\vertex[below right=of y] (d){\(\va_1\)};
				\vertex[above left=of z] (e){\(\va_1\)};
				\vertex[above right=of z] (f){\(\va_1\)};
				\vertex[below left=of k] (g){\(\va_1\)};
				\vertex[below right=of k] (h){\(\va_1\)};
							
				\diagram*{
					(z) --[scalar, quarter right] (x),
					(y) --[scalar, quarter right] (z),
					(k) --[scalar, quarter right] (y),
					(x) --[scalar, quarter right] (k),
					(x) --[scalar] (a),
					(x) --[scalar] (b),
					(z) --[scalar] (e),
					(z) --[scalar] (f),
					(k) --[scalar] (g),
					(k) --[scalar] (h),
					(y) --[scalar] (c),
					(y) --[scalar] (d),
				};
			\end{feynman}
		\end{tikzpicture}+...
		\caption{The 1-loop contributions of the scalar fields}\label{fig1}		
	\end{figure}

	Also the squared masses of the gauge and Higgs particles all can be split into two separate components at the tree level. From Eqs.~\eqref{mchi1a} and \eqref{mrho1}, averaging over space and using Bose-Einstein and Fermi-Dirac distributions respectively for bosons and fermions to average over space, the one-loop effective potential can be obtained at high temperatures. Also according to the analysis of Appendix \ref{multi}, the analysis of the Lagrangian  of 2HDM into two separate components (as shown in Secs. \ref{iiia}, \ref{iiib}, \ref{iiic}),  the total effective potential in the 2HDM model can be rewritten as
	\be\label{EP-E331}
		V^{2HDM}_{eff}= V_{eff}^{2HDM}(v_1)+V_{eff}^{2HDM}(v_2).
	\ee
	For further clarity, we restate the calculating process of the effective potential from the contributions of one-loop diagrams. The process of calculating the one-loop effective potential is the process of calculating the contribution of 1-loop diagrams with $n$ external lines that are Higgs scalar fields (fields that act as mass generators). In the 2HDM, there are two Higgs fields ($h, H$) that act as such, corresponding to two VEVs ($v_1, v_2$). One-loop diagrams are shown in Figs.~\ref{fig1}, \ref{fig2}, \ref{fig3}, \ref{fig4}.
	
	\begin{figure}[h!]
		\begin{tikzpicture}[baseline=(current  bounding  box.center)]
			\begin{feynman}
				\vertex (x);
				\vertex[right=1cm of x] (y);
				\vertex[left=of x] (a){\(\va_1\)};
				\vertex[right=of y] (d){\(\va_1\)};
				\vertex[above right= 0.5cm and 0.75cm of x] (p);
				
				\diagram*{
					(y) --[fermion, half right] (x),
					(x) --[fermion, half right] (y),
					(x) --[scalar] (a),
					(y) --[scalar] (d),
				};
			\end{feynman}
		\end{tikzpicture}
		+
		\begin{tikzpicture}[baseline=(current  bounding  box.center)]
			\begin{feynman}
				\vertex (x);
				\vertex[right=1.5cm of x] (y);
				\vertex[above right=1cm of x] (z);
				\vertex[below right=1cm of x] (k);
				\vertex[left=of x] (a){\(\va_1\)};
				\vertex[right=of y] (c){\(\va_1\)};
				\vertex[above=of z] (e){\(\va_1\)};
				\vertex[below=of k] (g){\(\va_1\)};
				
				\diagram*{
					(z) --[fermion, quarter right] (x),
					(y) --[fermion, quarter right] (z),
					(k) --[fermion, quarter right] (y),
					(x) --[fermion, quarter right] (k),
					(x) --[scalar] (a),
				
					(z) --[scalar] (e),
					
					(k) --[scalar] (g),
					
					(y) --[scalar] (c),
					};
			\end{feynman}
		\end{tikzpicture}
		+
		\begin{tikzpicture}[baseline=(current  bounding  box.center)]
			\begin{feynman}
				\vertex (x);
				\vertex[right=2cm of x] (y);
				\vertex[above right=1cm of x] (z);
				\vertex[below right=1cm of x] (k);
				\vertex[above left=1cm of y] (m);
				\vertex[below left=1cm of y] (n);
				\vertex[left=of x] (a){\(\va_1\)};
				\vertex[above left=of z] (b){\(\va_1\)};
				\vertex[right=of y] (c){\(\va_1\)};
				\vertex[above right=of m] (d){\(\va_1\)};
				\vertex[below left=of k] (e){\(\va_1\)};
				\vertex[below right=of n] (f){\(\va_1\)};
						
				\diagram*{
					(z) --[fermion, quarter right] (x),
					(m) --[fermion, quarter right,looseness=0.275] (z),
					(y) --[fermion, quarter right] (m),
					(n) --[fermion, quarter right] (y),
					(k) --[fermion, quarter right,looseness=0.275] (n),
					(x) --[fermion, quarter right] (k),
					(x) --[scalar] (a),
					(z) --[scalar] (b),
					(m) --[scalar] (d),
					(n) --[scalar] (f),
					(k) --[scalar] (e),
					(y) --[scalar] (c),
				};
			\end{feynman}
		\end{tikzpicture}+...
		\caption{The 1-loop contributions of the fermion fields}\label{fig2}
	\end{figure}
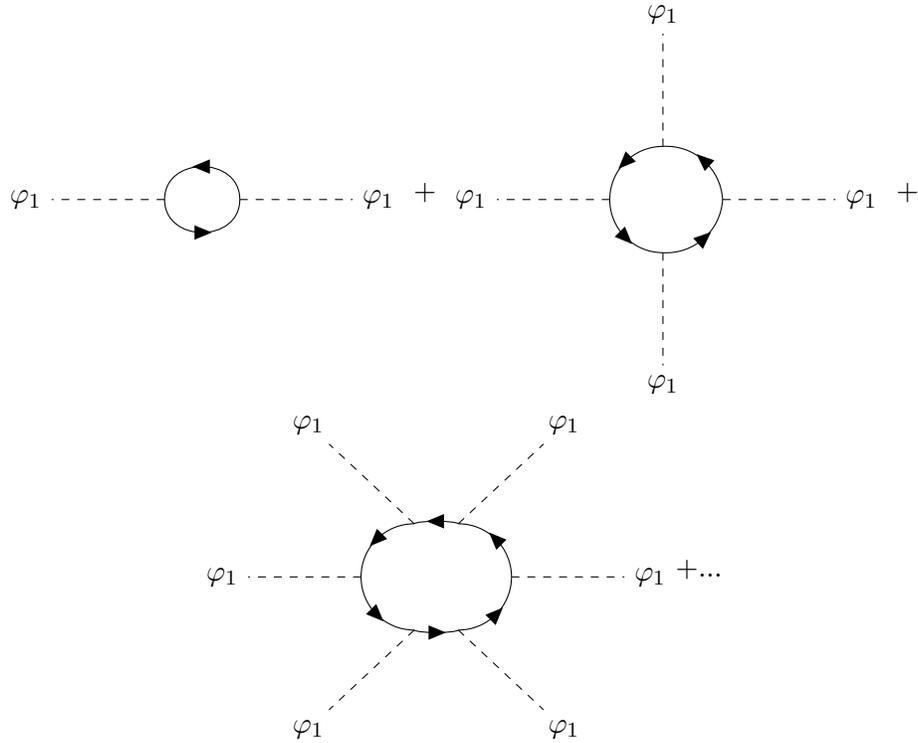	
	\begin{figure}[h]
		\begin{tikzpicture}[baseline=(current  bounding  box.center)]
			\begin{feynman}
				\vertex (x);
				\vertex[right=1.25cm of x] (y);
				\vertex[above left=of x] (a){\(\va_1\)};
				\vertex[below left=of x] (b){\(\va_1\)};
				\vertex[above right= 0.5cm and 0.75cm of x] (p);
				
				\diagram*{
					(y) --[boson, half right] (x),
					(x) --[boson, half right] (y),
					(x) --[scalar] (a),
					(x) --[scalar] (b),
				};
			\end{feynman}
		\end{tikzpicture}
		+
		\begin{tikzpicture}[baseline=(current  bounding  box.center)]
			\begin{feynman}
				\vertex (x);
				\vertex[right=1.5cm of x] (y);
				\vertex[above left=of x] (a){\(\va_1\)};
				\vertex[below left=of x] (b){\(\va_1\)};
				\vertex[above right=of y] (c){\(\va_1\)};
				\vertex[below right=of y] (d){\(\va_1\)};
				\diagram*{
					(y) --[boson, half right] (x),
					(x) --[boson, half right] (y),
					(x) --[scalar] (a),
					(x) --[scalar] (b),
					(y) --[scalar] (c),
					(y) --[scalar] (d),
				};
			\end{feynman}
		\end{tikzpicture}
		+
		\begin{tikzpicture}[baseline=(current  bounding  box.center)]
			\begin{feynman}
				\vertex (x);
				\vertex[right=2cm of x] (y);
				\vertex[above right=1.25cm of x] (z);
				\vertex[below right=1.25cm of x] (k);
				\vertex[above left=of x] (a){\(\va_1\)};
				\vertex[below left=of x] (b){\(\va_1\)};
				\vertex[above right=of y] (c){\(\va_1\)};
				\vertex[below right=of y] (d){\(\va_1\)};
				\vertex[above left=of z] (e){\(\va_1\)};
				\vertex[above right=of z] (f){\(\va_1\)};
				\vertex[below left=of k] (g){\(\va_1\)};
				\vertex[below right=of k] (h){\(\va_1\)};
						
				\diagram*{
					(z) --[boson, quarter right] (x),
					(y) --[boson, quarter right] (z),
					(k) --[boson, quarter right] (y),
					(x) --[boson, quarter right] (k),
					(x) --[scalar] (a),
					(x) --[scalar] (b),
					(z) --[scalar] (e),
					(z) --[scalar] (f),
					(k) --[scalar] (g),
					(k) --[scalar] (h),
					(y) --[scalar] (c),
					(y) --[scalar] (d),
				};
			\end{feynman}
		\end{tikzpicture}
		+...
		\caption{The 1-loop contributions of the gauge fields}\label{fig3}
	\end{figure}
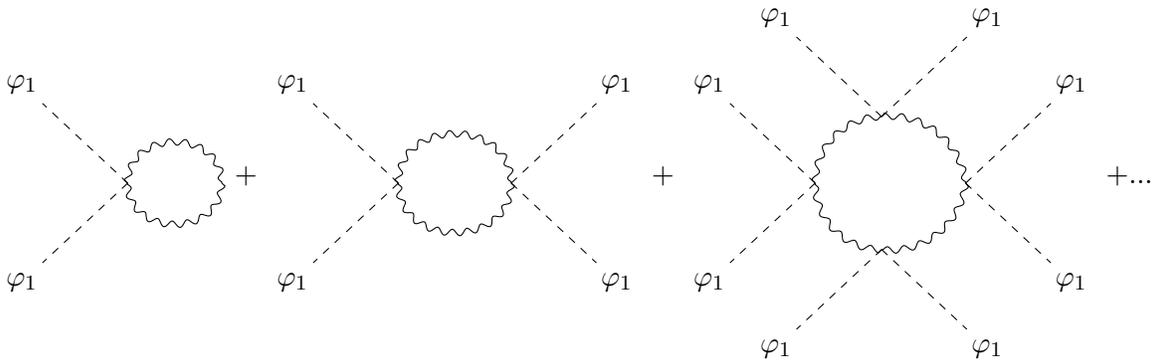
	
	Similarly, also having diagrams where the external lines are $\va_2$.  The mixing $\va_1-\va_2$ diagrams as Fig.~\ref{fig4}.
	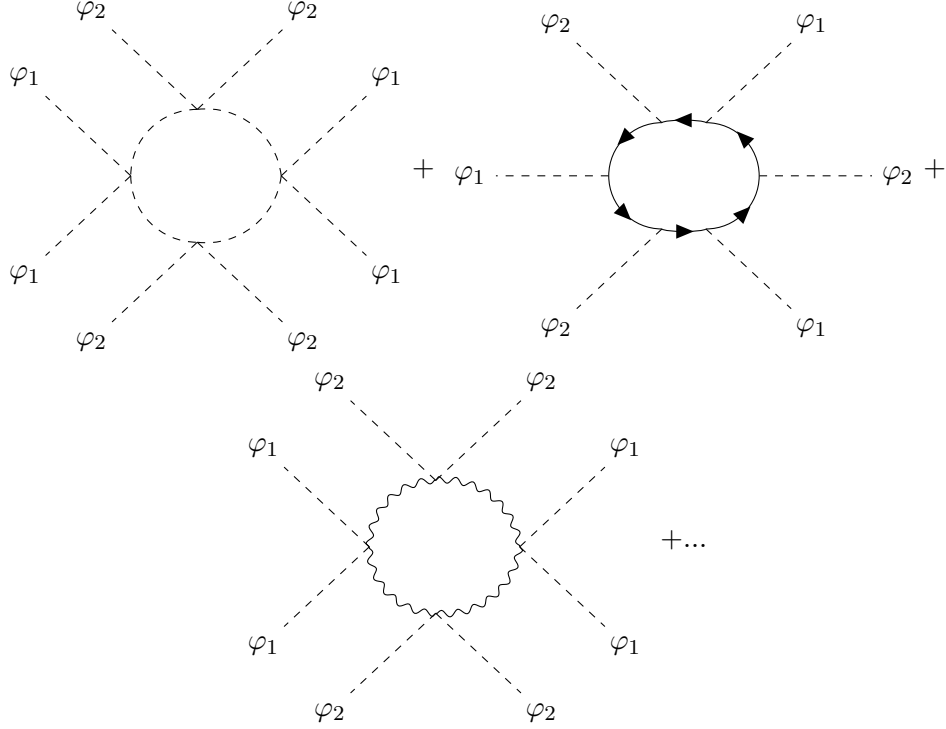
\begin{figure}[h]
		\begin{tikzpicture}[baseline=(current  bounding  box.center)]
			\begin{feynman}
				\vertex (x);
				\vertex[right=2cm of x] (y);
				\vertex[above right=1.25cm of x] (z);
				\vertex[below right=1.25cm of x] (k);
				\vertex[above left=of x] (a){\(\va_1\)};
				\vertex[below left=of x] (b){\(\va_1\)};
				\vertex[above right=of y] (c){\(\va_1\)};
				\vertex[below right=of y] (d){\(\va_1\)};
				\vertex[above left=of z] (e){\(\va_2\)};
				\vertex[above right=of z] (f){\(\va_2\)};
				\vertex[below left=of k] (g){\(\va_2\)};
				\vertex[below right=of k] (h){\(\va_2\)};
							
				\diagram*{
					(z) --[scalar, quarter right] (x),
					(y) --[scalar, quarter right] (z),
					(k) --[scalar, quarter right] (y),
					(x) --[scalar, quarter right] (k),
					(x) --[scalar] (a),
					(x) --[scalar] (b),
					(z) --[scalar] (e),
					(z) --[scalar] (f),
					(k) --[scalar] (g),
					(k) --[scalar] (h),
					(y) --[scalar] (c),
					(y) --[scalar] (d),
				};
			\end{feynman}
		\end{tikzpicture}+
		\begin{tikzpicture}[baseline=(current  bounding  box.center)]
			\begin{feynman}
				\vertex (x);
				\vertex[right=2cm of x] (y);
				\vertex[above right=1cm of x] (z);
				\vertex[below right=1cm of x] (k);
				\vertex[above left=1cm of y] (m);
				\vertex[below left=1cm of y] (n);
				\vertex[left=of x] (a){\(\va_1\)};
				\vertex[above left=of z] (b){\(\va_2\)};
				\vertex[right=of y] (c){\(\va_2\)};
				\vertex[above right=of m] (d){\(\va_1\)};
				\vertex[below left=of k] (e){\(\va_2\)};
				\vertex[below right=of n] (f){\(\va_1\)};
								
				\diagram*{
					(z) --[fermion, quarter right] (x),
					(m) --[fermion, quarter right,looseness=0.275] (z),
					(y) --[fermion, quarter right] (m),
					(n) --[fermion, quarter right] (y),
					(k) --[fermion, quarter right,looseness=0.275] (n),
					(x) --[fermion, quarter right] (k),
					(x) --[scalar] (a),
					(z) --[scalar] (b),
					(m) --[scalar] (d),
					(n) --[scalar] (f),
					(k) --[scalar] (e),
					(y) --[scalar] (c),
				};
			\end{feynman}
		\end{tikzpicture}+
		\begin{tikzpicture}[baseline=(current  bounding  box.center)]
			\begin{feynman}
				\vertex (x);
				\vertex[right=2cm of x] (y);
				\vertex[above right=1.25cm of x] (z);
				\vertex[below right=1.25cm of x] (k);
				\vertex[above left=of x] (a){\(\va_1\)};
				\vertex[below left=of x] (b){\(\va_1\)};
				\vertex[above right=of y] (c){\(\va_1\)};
				\vertex[below right=of y] (d){\(\va_1\)};
				\vertex[above left=of z] (e){\(\va_2\)};
				\vertex[above right=of z] (f){\(\va_2\)};
				\vertex[below left=of k] (g){\(\va_2\)};
				\vertex[below right=of k] (h){\(\va_2\)};
								
				\diagram*{
					(z) --[boson, quarter right] (x),
					(y) --[boson, quarter right] (z),
					(k) --[boson, quarter right] (y),
					(x) --[boson, quarter right] (k),
					(x) --[scalar] (a),
					(x) --[scalar] (b),
					(z) --[scalar] (e),
					(z) --[scalar] (f),
					(k) --[scalar] (g),
					(k) --[scalar] (h),
					(y) --[scalar] (c),
					(y) --[scalar] (d),
				};
			\end{feynman}
		\end{tikzpicture}
		+...
		\caption{The diagrams of one-loop contribution when there is a mix of $v_1-v_2$}\label{fig4}
	\end{figure}

In the above diagrams, $\va_1$ and $\va_2$ correspond to the terms of Higgs fields that only contain $v_1$ or $v_2$. Since the 2HDM consists of two VEVs, having the mixing diagrams as in Fig.~\ref{fig4}.	
	
For the $\phi^4$ theory, calculating the contributions of the diagrams as in Figs.~\ref{fig1},\ref{fig2} and \ref{fig3} is really easy, and this has been shown in Ref.~\cite{quiros}. However, the difficulty is to calculate the contributions of the diagrams in Fig.~\ref{fig4}. As in Appendix \ref{multi}, we explicitly study the first diagram in Fig.~\ref{fig4} that translates to the following expression:
	
	\begin{align}
		\Ga_n(v_1,v_2)=&i\fr{1}{2n}\int \fr{d^4p}{(2\pi)^4}\left(\fr{i}{p^2-m^2(v_1,v_2)+i\varep}\right)^n (-i 2\la_i)^{n_1/2}v_1^{n_1}(-i2\la_2)^{n_2/2}v_2^{n_2},
	\end{align}	
with $n_1+n_2=2n$. Summing over all $\Ga_n$ with $n$ runs from $0$ to infinity, and hence $n_1$ and $n_2$ will also run from $0$ to infinity. Hence calculating the integrals $\Ga_n$ and the infinite sums are very tricky. Since for each value of $n$, there are a sum that runs with $n_1$ or $n_2$.	
	
To quickly calculate $\Ga_n$, from remarks 2 and 4, $\Ga_n(v_1,v_2)\equiv \Ga_n(v_1)$,
	\begin{align}
		\Ga_n(v_1,v_2)=&i\fr{1}{2n}\int \fr{d^4p}{(2\pi)^4}\left(\fr{i}{p^2-m^2(v_1)+i\varep}\right)^nv_1^{n_1}v_1^{n_2}\kappa^{n_2}(-i 2 \la_1)^{n_1/2}(-i 2 \la_2)^{n_2/2}\\
		&=i\fr{1}{2n}\int \fr{d^4p}{(2\pi)^4}\left(-i 2 \la'_n \fr{i}{p^2-m^2(v_1)+i\varep}\right)^nv_1^{2n}.
	\end{align}
Here, $\la'_n$ must have some very small values in order for the series to converge. Therefore $\la'_n\sim \la'$. Hence, $\Ga_n(v_1,v_2)\equiv \Ga_n(v_1)$,
	\begin{align}
		\Ga_n(v_1)=i\fr{1}{2n}\int \fr{d^4p}{(2\pi)^4}\left( 2\la'\fr{v^2_1}{p^2-m^2(v_1)+i\varep}\right)^n.
	\end{align}

Therefore the contributions from these mixing diagrams can be combined into the contribution from the diagrams that the external lines are just all $\va_1$. In other words, by calculating the contributions from the diagrams, and applying Remarks 2, and 4, the mixing diagrams can be processed to turn the effective potential into two clearly separated components, one depends only on $v_1$, while the other depends only on $v_2$. In other words, the effective potential will be expressed as Eq.~(\ref{EP-E331}).
	
In the previous studies, to overcome the difficulties of dealing with the mixing terms of VEVs and to investigate the process of electroweak symmetry breaking similar to the SM, the authors have converted $v_1$ and $v_2$ to $v$ through $\tan\beta$ \cite{tdhm,Fuyutob}. This is a very clever technique, but in terms of physics, it needs to be interpreted with care. Since $v_1\ne v_2$, the symmetry-breaking stages must be sequential. The fact that we write the same effective potential for $v$ (shortly denoted as $V_{eff}(v)$) to calculate the strength of phase transition is not wrong, but we should only write it when $v_1\ll v_2$, to clearly show the nature of the physics in the 2HDM. The limitations and utilities of this technique will be analyzed in the following sections where investigating the 2HDM-$S_3$.

\subsection{Comments on EWPT in the 2HDM}\label{IIE}

First, we have summarized  of experimental as well as theoretical calculations leading to parameter regions in 2HDM models: 
\begin{itemize}
		
	\item Since the 125 GeV Higgs boson observed at the LHC, the model becomes consistent with the LHC Higgs data when the model provides such a Higgs particle \cite{bpal1}. $1<\tan\beta$ can be precisely determined from the requirement of the light mass of the up- and down-quarks \cite{bpal1}. From here also admit the scenario $\mu_1, \mu_2, l_1, l_2 \neq 0$ and having the soft-breaking potential. Also according to Ref.~\cite{bpal2}, the mass of the nonstandard particles are less than 1 TeV, $0.3<\tan\beta<17$.
	
	\item Some references used the data from decay channels in LHC and investigated the value range of $\tan\beta$. In Ref. \cite{beta1}, the authors removed the circumstance $\tan\beta < 1$ for four types of 2HDM. In Ref. \cite{beta2}, the authors saw that in the decay channel $pp \rightarrow W \rightarrow h H^\pm$, the scattering amplitude of that channel is lasting over the change of $\tan\beta$ in the range of $1 < \tan\beta <10$.
		
	\item The FCNC structure 
	exists in the model. It is therefore compatible with current experimental data on quarks \cite{bpal1,36t}. The mass of exotic particle below 190 or 300 GeV has been already excluded by the data from the LHC Run-II and the HL-LHC, the most of the parameter region would be explored \cite{mayumi}. Since having a significant amount of the $b\bar{b}$ branching ratios for the additional Higgs bosons \cite{mayumi}.
	
	\item The model under consideration contains the pseudoscalar field $A$ being attached subject for recent experimental study \cite{CMS2}.
	
	\item Next, in general, there are four types of 2HDM models of $Z_2$ symmetry. Two of the four types were investigated in Ref.~\cite{dori} and also with parameter domains consistent with the above 
	conclusions.
\end{itemize}

The above conclusions lead to an instruction for surveying 2HDM-$S_3$ also with a parameter region, $1\sim\tan\beta<17$ and the masses of additional bosons must be larger than 200 GeV.\\

More importantly, the following comments about the EWPT in 2HDM models after the observation of  the 125 GeV Higgs  at LHC (2012 are in order):
\begin{itemize}
	\item According to Ref. \cite{1305.6610}, with LHC data and decay channel $h^0\longrightarrow \gamma\gamma$, for a strong first order EWPT in 2HDM, $m_A>400$ GeV, a mass hierarchy $m_{H^\pm }<m_{H} < m_{A}$ and $1 < \tan\beta <10$. This does not define the upper bounds of the masses of the particles, but the lower bounds are about 400 GeV.
	
	\item In Ref.~\cite{1504.05949}, when analyzing the inert 2HDM model, for $500 \text{ GeV}<m_{H}<a few$ TeV, for a first order EWPT, boson $H$ could be a candidate for dark matter.
	
	\item The authors in Ref.~\cite{PRL} performed a nonperturbation study of EWPT in 2HDM. To have a first order EWPT, the condition is $m_{A}>m_{H}+m_{Z}$. This suggests that the mass of additional Higgs bosons must be larger mass than one of SM-like Higgs boson.
	
	\item In Ref.~\cite{2111.13079J. High Energy Phys}, for a first order EWPT and combined with LHC data, the masses of additional Higgs bosons are typically $300-400$ GeV. The triple Higgs boson coupling is predicted to be $35-55$ \% larger than the standard model value.
	
	\item In Ref.~\cite{dori}, analyzing decay channels $A, H, H^{\pm}\longrightarrow tt, tb$, combining with HL-LHC signal and gravitational wave observations at LISA. The 2HDM model for a first order EWPT. It also shows that these decay channels can be key channels to authenticate the first order EWPT in 2HDM.
	
	\item In particular, in Ref.~\cite{dori}, Fig. 1 shows that the ratio between two phase transition strengths of the two model types is almost independent of $\tan\beta$. Different scenarios between the masses of the additional Higgs particles were analyzed in the EWPT problem, such as $m_{H^\pm}=m_A$ or $m_{H}=m_{H^\pm }$.
\end{itemize}

The summaries of the EWPT results in the 2HDM model are important indicators for the parameter space in the calculation of EWPT in the 2HDM-$S_3$. It also shows the effect of the $S_3$ symmetry, which will be discussed in the following sections.

\section{Review on the 2HDM-$S_3$}\label{iv}
\subsection{Particle content}

To solve with FCNC for the 2HDM, ones  can realize by implement of $S_3$ symmetry \cite{S32dM}. The particle contents and their  charge assignment are given in Table \ref{tab:1}.

	\begin{table}[h]
\begin{tabular}{|c|c|c|c|c|c|c|c|c|c|}
\hline Particle &$Q_a$& $L_{a}$ & $L_\tau$ &$u_{aR}$&$d_{iR}$& $ e_{aR} $ & $ \tau_R $  
& $\Phi_1 $& $\Phi_2 $  \\
\hline
$SU(2)_L$& $\bm{2}$ & $\bm{2}$ & $\bm{2}$ &$\bm{1}$&$1$& $\bm{1}$& $\bm{1}$ &  $\bm{2}$
& $\bm{2}$    
 \\
\hline
$U(1)_Y$& $\fr 1 6$ & $- \fr 1 2$ & $- \fr 1 2$ &$\fr 2 3$&$- \fr 1 3$& $-1$& $-1$ &  $\fr 1 2$
& $\fr 1 2$    \\\hline
$\mathbb{S}_3$ & $\mathbf{1}$ & $\mathbf{2}$ & $\mathbf{1}$ & $\mathbf{1'}$ &$\mathbf{1}$ or $\mathbf{1'}$& $\mathbf{2}$& $\mathbf{1}$ or $\mathbf{1'}$ & $\mathbf{1}$&  $\mathbf{1'}$ \\
\hline
\end{tabular}%
\caption{The particle contents and their charge assignment of the $SU(2)_L\times U(1)_Y\times S_3$ symmetry. }
\label{tab:1}
\end{table}

There are two kinds of representations for $S_3$: real and complex; and it is easier to work with 
complex representation \cite{ma}.  

	\subsection{Higgs potential}
	
	The generic scalar potential of 2HDM-$S_3$ \cite{1601} can be written as
	\begin{align}
		\begin{split}
			\label{vh}
			V_H = \quad &m_{11}^2 \Phi_1^2 + m_{22}^2 \Phi_2^2 - m_{12}^2 \Phi_1. \Phi_2
			- (m_{12}^2)^* \Phi_2 . \Phi_1\\
			& +\dfrac{\la_1}{2} \Phi_1^4 + \dfrac{\la_2}{2} \Phi_2^4 + \la_3 \Phi_1^2 \Phi_2^2 +\la_4 (\Phi_1 . \Phi_2)( \Phi_2 . \Phi_1)\\
			&+ \left[ \dfrac{\la_5}{2} (\Phi_1 . \Phi_2)^2 + (\la_6 \Phi_1^2 + \la_7 \Phi_2^2)(\Phi_1 . \Phi_2) + \textrm{H. c.} \right],
		\end{split}
	\end{align}
	where any couplings other than $m^2_{12}$, $\la_5$, $\la_6$, and $\la_7$ are real. Using $v_1/\sqrt{2}$ and $v_2/\sqrt{2}$ to denote the VEVs, but ignore the factor $\sqrt{2}$ (however, it was still included in the calculations).

	If choosing $\Phi \sim (\mathbf{1},\mathbf{1'})$ to be the representations of $S_3$, then all the odd terms only containing $\Phi_2$, such as $ m_{12}^2 \Phi_1 . \Phi_2$, $(m_{12}^2)^* \Phi_2 . \Phi_1$ and $(\la_6 \Phi_1^2 + \la_7 \Phi_2^2)(\Phi_1 . \Phi_2) $ must be terminated for  the Lagrangian to be invariant under the $S_3$ group transformation, since the representation $\mathbf{1'}$ changes the sign of the fields with odd permutation. In this case, getting a $Z_2$ symmetric potential, in which $m^2_{12} = \la_6 = \la_7=0$.
	
	In general, we can assume that $\Phi \sim s$, where $s$ denotes either of the two alternatives $(\mathbf{1},\mathbf{1})$ or $(\mathbf{1},\mathbf{1'})$. Because the labels 1 and 2 were selected at random, the case $\Phi \sim (\mathbf{1'},\mathbf{1})$ is also included. For its simplicity, a complicated representation can be selected to work with. The most common Higgs potential of a $S_3$ doublet is of the following form where the two scalars $\Phi = (\phi_1, \phi_2)^T$ transform as a doublet in a \textit{complex} representation \cite{1601}:
		
	\begin{align}
		\begin{split}
			V_C= \quad &\mu_1^2 (\phi_2^2 + \phi_1^2) + \fr{1}{2} l_1 (\phi_2^2 + \phi_1^2)^2 + \fr{1}{2} l_2 (\phi_2^2 - \phi_1^2)^2 \\
			&+ l_3 (\phi_1 .  \phi_2) (\phi_2 .  \phi_1). \label{vC}
		\end{split}
	\end{align}
Note that both complex and real representations provide the similar  result  given  in \eqref{vC} which coincides with generic one  $V_H$ in \eqref{vh} for
the conditions \cite{1601}
\be
 m^2_{11} = m^2_{22} \equiv \mu^2_1\,,  \, m^2_{12} = 0\,, \la_1 = \la_2 \,, \,  \la_5
 = 0 \,.
 \label{l2}
 \ee

	\subsection{The soft breaking of $S_3$ group}
	
	To break $S_3$ softly, Ma and Melic \cite{melic} include a soft potential by hand to the full one, while still preserving the $\phi_1 \leftrightarrow \phi_2$ symmetry,
	\begin{align}
		V_{\text{soft}} = - \mu_2^2 \left( \phi_1^\+ \phi_2 + \phi_2^\+ \phi_1  \right).
	\end{align}
Soft breaking terms here mean that they violate the original symmetry in the Higgs potential or the Lagrangian. They are 'soft' because the couplings associated with those terms are small. Without the above mentioned term, we will face the trouble of massless pseudoscalar $A$. Realistically, there should be some breaking terms which will take care of this problem.
	
	In a spontaneously broken case, you break the symmetry of the ground state and it naturally breaks the symmetry in the Lagrangian. On the other hand, in this case, those terms must be added by hand to break the symmetry. The potential then becomes $ V_C + V_{\text{soft}} $.	
	
	The term containing $V_{soft}$ makes the mass of the pseudoscalar Higgs particle $A$ always nonzero in all cases. If there is no $V_{soft}$ term, the Higgs particle A will have a mass of 0, when $v_1=v_2$. Besides, the term $V_{soft}$ must exist, since it represents the direct interaction between $\phi_1$ and $\phi_2$ Therefore, we must study the Higgs potential that contains the term $V_{soft}$.

	\section{Electroweak phase transition in the 2HDM-$S_3$}\label{v}
	\subsection{A vital role of $S_3$}
	
	First, the function of the $S_3$ group in the 2HDM-$S_3$ can be shown by comparing the Higgs potential of the two models before and after adding $S_3$ symmetry. Let us  consider the Higgs potential of 2HDM which has the following form
	\begin{align}
	V_\text{Higgs}^{\text{2HDM}} & =  m_{11}^2 \Phi_1^\+ \Phi_1+ m_{22}^2 \Phi_2^\+ \Phi_2 - m_{12}^2 \left( \Phi_1^\+ \Phi_2 + \Phi_2^\+ \Phi_1 \right) + \dfrac{\la_1}{2} \left(  \Phi_1^\+ \Phi_1 \right)^2\label{l4}	\\
		&+ \dfrac{\la_2}{2} \left(  \Phi_2^\+ \Phi_2 \right)^2
		+\la_3 \Phi_1^\+ \Phi_1 \Phi_1^\+ \Phi_2 + \la_4 \Phi_1^\+ \Phi_2 \Phi_2^\+ \Phi_1 + \dfrac{\la_5}{2} \left[   \left(  \Phi_1^\+ \Phi_1 \right)^2 + \left(  \Phi_2^\+ \Phi_2 \right)^2  \right].\nn
	\end{align}

	The above Higgs potential  contains 8 parameters, and  the squared masses of Higgs particles are given by
	\begin{align}
		\begin{split}
			& m_{H^\pm,2HDM}^2  = \left( \dfrac{m_{12}^2}{v_1 v_2} - \la_4 -
			 \la_5  \right) v^2,\\
			& m_{A,2HDM}^2 = \left( \dfrac{m_{12}^2}{v_1 v_2} - 2\la_5 \right) (v_1^2 
			+ v_2^2),\\
			& m_{h,2HDM}^2 = \dfrac{1}{2} \left[ (\la_1 v_1^2 + \la_2 v_2^2) + m_{12}^2 \dfrac{v^2}{v_1 v_2} \right]\label{l3}\\
			&\qquad - \sqrt{\left[ \dfrac{\la_1 v_1^2 - \la_2 v_2^2}{2} - 
			\dfrac{m_{12}^2}{2 v_1 v_2} (v_1^2 - v_2^2)\right]^2 + (\la_{345} v_1 v_2 - m_{12}^2)^2},\\
			&m_{H,2HDM}^2 = \dfrac{1}{2} \left[ (\la_1 v_1^2 + \la_2 v_2^2) + m_{12}^2 \dfrac{v^2}{v_1 v_2} \right] \\
			&\qquad + \sqrt{\left[ \dfrac{\la_1 v_1^2 - \la_2 v_2^2}{2} - 
			\dfrac{m_{12}^2}{2 v_1 v_2} (v_1^2 - v_2^2)\right]^2 + (\la_{345} v_1 v_2 - m_{12}^2)^2}.
		\end{split}
	\end{align}
	
 Looking at  the  formulas of masses in \eqref{l3}, we see that they contain the very annoying mixing terms of $v_1$ and $v_2$. Hence, at nonzero temperatures, these mixing terms make the job of calculating the contributions from the particles to the effective potential very difficult. They turn the symmetry breaking process into the process of breaking the ambiguous mixings of $\Phi_1$ and $\Phi_2$.
	
Meanwhile, in the complex representation, the Higgs potential of 2HDM-$S_3$ has a simpler form:
	\begin{align}	
		V_\text{Higgs}^{\text{2HDM} \otimes S_3} \equiv & \, V(\phi_1, \phi_2)\crn
		 & = \mu_1^2 (\phi_2^2 + \phi_1^2) + \fr{1}{2} l_1 (\phi_2^2 + \phi_1^2)^2 + \fr{1}{2} l_2 (\phi_2^2 - \phi_1^2)^2 \crn
		&\qquad+ l_3 (\phi_1 .  \phi_2) (\phi_2 .  \phi_1) - \mu_2^2 \left( \phi_1 . \phi_2 + \phi_2 .  \phi_1  \right).
	\end{align}
	
	Seeing the structure of those Higgs potentials containing  $S_3$ symmetry, ones conclude that  the Higgs potential becomes simpler, with fewer parameters (from 8 reduced to 5). Thanks to that, the forms of Higgs mass are also simpler. By replacing $v_1$ and $v_2$ by $c_\beta v$ and $s_\beta v$, in expressions of $H$ and $h$ masses, ones get  a form of $v$-dependent. 
	\begin{align}
		\begin{split}
			&m_{H^\pm,S_3}^2 =  -l_2 v^2,\\
			&m_{A,S_3}^2 = - \dfrac{1}{2} (2l_2 - l_3) v^2 ,\\
			& m_{H,S_3}^2 = \dfrac{1}{4} v^2
			\bigg[ 2 l_1 + l_3  + \sqrt{ 16 l_1 l_2 (c_\beta^2 - s_\beta^2)^2 - 8 l_2 l_3  + l_3^2 - 4 l_1 l_3 (c_\beta^4 - 6 c_\beta^2 s_\beta^2 + s_\beta^4) } \bigg]=\fr{1}{4}f_H.v^2,\label{blkl}\\
			& m_{h,S_3}^2 = \dfrac{1}{4} v^2
			\bigg[ 2 l_1 + l_3  - \sqrt{ 16 l_1 l_2 (c_\beta^2 - s_\beta^2)^2 - 8 l_2 l_3  + l_3^2 - 4 l_1 l_3 (c_\beta^4 - 6 c_\beta^2 s_\beta^2 + s_\beta^4) } \bigg]=\fr{1}{4}f_h.v^2\,.
		\end{split}
	\end{align}

	The $S_3$ symmetry has removed the mixing parts of the two VEVs inside the mass of two charged Higgs bosons $H^\pm$ and one neutral Higgs boson $A$. Otherwise, $h$ and $H$ still have the mixing parts hiding in $c_\beta$ and $s_\beta$. However, this mixing would be simpler in the 2HDM, since there are no such ratios as $v_1/v_2$ or $v^2/(v_1.v_2)$, their mass formulas then have fewer parameters and are simpler than the ones of 2HDM without $S_3$.
	
	Therefore, according to the remarks for the 2HDM, $S_3$ had made the process of the electroweak phase transition occur in each VEV. This will be discussed in the next section.

	\subsection{Structure of EWPT}\label{stewpt}
	The procedure to describe the structure of electroweak phase transition in this model is similar to that in the SM, whereas Higgs and gauge bosons are the main contributors in the breaking symmetry process. For that reason, determination mass can also affect the phase transition.
	
	The 2HDM with $S_3$ symmetry has Higgs Lagrangian with kinetic and potential elements as
	\begin{align}
		\mathcal{L}_{\text{Higgs}} = ( D_\mu \phi_1)^\+ (D^\mu \phi_1 ) + ( D_\mu \phi_2)^\+ (D^\mu \phi_2 ) - V(\phi_1, \phi_2)\, .
	\end{align}
	Averaging all over the space, then replacing fields with VEVs, the Higgs Lagrangian with variables $v_1$ and $v_2$ (with $v_1 \neq v_2$) has the following form:
	\begin{align}
		\begin{split}
			\mathcal{L}_{\text{Higgs}} &= \dfrac{1}{2} \pa^\mu v_1 \pa_\mu v_1 + \dfrac{1}{2} \pa^\mu v_2 \pa_\mu v_2 - V_0 (v_1, v_2) \\
			& \qquad \sum_{i=\text{vector boson}} m_i^2 (v_1, v_2) W^\mu W_\mu + \sum_{j=\text{Higgs boson}} m_j^2 (v_1, v_2) H^2,
		\end{split}
	\end{align}
	whereas $W$ and $H$ are the vector boson and scalar fields,  respectively.
	
	Table \ref{mass1} contains the squared mass of the particles contributing to the EWPT, in the form of depending VEVs; $n$ is the degree of freedom of the fields. The masses of known particles are in generic form and at 0K, shown in Table \ref{mass2}.
	\begin{table}[htp]
		\begin{tabular}{ c|c|c|c|c }
			\hline
			Particles & $m^2(v_1,v_2)$ & $m^2(v_1)$ & $m^2(v_2)$ & $n$ \\
			\hline
			$m^2_{W^\pm}$ & $\fr{g^2v^2}{4}$ &$\fr{g^2v_1^2}{4}$ &$\fr{g^2v_2^2}{4}$  & $6 $\\
			
			$m^2_Z$ & $(g^2+ g'^2)\fr{v^2}{4}$ & $(g^2+ g'^2)\fr{v_1^2}{4}$ & $(g^2+ g'^2)\fr{v_2^2}{4}$ & $3$ \\
			
			$m^2_h$ & $ \fr{1}{4} f_h v^2$ &$ \fr{1}{4} f_h v_1^2 $& $\fr{1}{4} f_h v_2^2$ & $1$ \\
			
			$m^2_H$ & $ \fr{1}{4} f_H v^2$ &$ \fr{1}{4} f_H v_1^2 $& $\fr{1}{4} f_H v_2^2$ & $1$ \\
			
			$m^2_A$ & $ - \dfrac{1}{2} (2l_2 - l_3) v^2  $ & $ - \dfrac{1}{2} (2l_2 - l_3) v_1^2  $ & $ - \dfrac{1}{2} (2l_2 - l_3) v_2^2  $ & $1$ \\
			
			$m^2_{H^\pm}$ & $ - l_2 v^2$ & $- l_2 v_1^2$ & $ - l_2 v_2^2 $ & $2$ \\
			
			$m_t^2$ & $f_t^2 v^2$ &$f_t^2 v_1^2$ & $f_t^2 v_2^2$ & $-12$ \\
			\hline
		\end{tabular}	
\centering
		\caption{Squared mass of the gauge bosons and scalar bosons in 2HDM-$S_3$; whereas mass of the $W^\pm$, $Z$ and $t$ is the same as the one in SM;  $v^2 = v_1^2 + v_2^2$.}	
		\label{mass1}
	\end{table}
	
	\begin{table}[htp]
		\centering
		\begin{tabular}{ c|c|c|c|c }
			\hline
			Particles & $m_{W^\pm}(v_0)$ & $m_{Z}(v_0)$  & $m_h(v_0)$ & $m_t(v_0)$ \\
			\hline
			$m(v_0)~[\textrm{GeV}]$ &  $80.442$ &$91.18$  &$125$ &$173.1$ \\
			\hline
		\end{tabular}
		\caption{Mass of particles (GeV) at $0$K  in 2HDM-$S_3$}
		\label{mass2}
		
	\end{table}
	
Table \ref{mass1} shows us that all the particles in the model depend on two VEVs. But $v_1$ and $v_2$ depend on each other, $v_2/v_1 = \tan\beta $. The forms of mass could be changed into one-VEV-depended ($ v = \sqrt{v_1^2 + v_2^2}$) by replacing $v_1 = v c_\beta$ and $v_2 = v s_\beta$. Therefore, in this model, assuming the remark 4, the electroweak phase transition can be considered as a dual transition, with two VEVs accomplished to the condition $ v = \sqrt{v_1^2 + v_2^2}$ and at 0K, $ v_0 = \sqrt{v_{01}^2 + v_{02}^2} = 246$ GeV.
	
Here in remark 4, it is also a note that the coefficient $\ka$ is now equal to $\tan\beta$. Carefully observing Table \ref{mass1}, although the masses of $h$ and $H$ can be split, there still exists the coefficient $\ka$ in $f_h$ and $f_H$. These constants are only meaningful in that the masses of $h$ and $H$ can be split into terms that each one of them depends only  on one VEV. In other words, the contributions of the mixing of $v_1$ and $v_2$ are all brought back to only one VEV is $v_1$ or $v_2$, and the difference between $v_1$ and $v_2$ is put into the constants $f_h$ and $f_H$. Hence, the investigation of phase transition is somewhat relatively easier. But in the end, the replacement is really not that important, since in the next sections, this coefficient in fact will not have any effects on the strength of phase transition.
	
\subsection{The effective potential}
	
This dual-phase transition has the participation of new particles as two charged Higgs $H^\pm$, one neutral \textit{CP}-odd $A$, and one neutral \textit{CP}-even $H$. More importantly, there are also the presence of SM particles as one neutral \textit{CP}-even Higgs boson $h$, two charged gauge bosons $W^\pm$, one neutral boson $Z$ and top quark $t$.
	
The effective potential for each stage can be calculated in two ways. The effective potentials only contain the contributions from the particles outside of the SM and the gauge bosons, SM-like Higgs boson and top quark. The other particles have small values of mass so they can be just ignored. The process of calculating the effective potential is in detail given in Ref.~\cite{pkll}.
	
The effective potential of one phase transition without daisy loops has the form:
	\begin{align}
		\begin{split}
			\fontsize{11pt}{15pt}\selectfont
			V_{eff}(\mathcal{V}, T) = &V_0(\mathcal{V}) + \dfrac{1}{64\pi^2} \bigg[  6m_{W^\pm}^4(\mathcal{V})\ln\dfrac{m_{W^\pm}^2(\mathcal{V})}{Q^2} +3 m_{Z}^4(\mathcal{V})\ln\dfrac{m_{Z}^2(\mathcal{V})}{Q^2} + 2m_{H^\pm}^4(\mathcal{V})\ln\dfrac{m_{H^\pm}^2(\mathcal{V})}{Q^2}\\
			&+m_{h}^4(\mathcal{V})\ln\dfrac{m_{h}^2(\mathcal{V})}{Q^2} + m_{H}^4(\mathcal{V})\ln\dfrac{m_{H}^2(\mathcal{V})}{Q^2}+ m_{A}^4(\mathcal{V})\ln\dfrac{m_{A}^2(\mathcal{V})}{Q^2} - 12m_{t}^4(\mathcal{V})\ln\dfrac{m_{t}^2(\mathcal{V})}{Q^2} \bigg]\\
			&+ \dfrac{T^4}{4\pi^2} \bigg[ 6F_-\left( \dfrac{m_{W^\pm}(\mathcal{V})}{T} \right) + 3F_-\left( \dfrac{m_{Z}(\mathcal{V})}{T} \right) + 2F_-\left( \dfrac{m_{H^\pm}(\mathcal{V})}{T} \right) + F_-\left( \dfrac{m_{h}(\mathcal{V})}{T} \right) \\
			&+ F_-\left( \dfrac{m_{H}(\mathcal{V})}{T} \right) + F_-\left( \dfrac{m_{A}(\mathcal{V})}{T} \right) +12F_+\left( \dfrac{m_{t}(\mathcal{V})}{T} \right) \bigg],
		\end{split}
	\end{align}
whereas
	\begin{align}
		&F_\pm\left(\dfrac{m_\phi}{T}\right)=\int_0^{\fr{m_\phi}{T}} \al J_\pm^{(1)}(\al,0)d\al,\\
		& J_{\pm}^{(1)}(\al,0)= 2\int_\al^{\infty} \dfrac{(x^2-\al^2)^{\nu/2}}{e^{x} \pm 1}dx.
	\end{align}
Then,
	\begin{align}
		&\begin{cases}
			&J_{-}^{(1)}(\al,0)= \dfrac{\pi^2}{3}-\pi\al- \dfrac{\al^2}{2}\left(\ln\dfrac{\al}{4\pi}+C-\dfrac{1}{2}\right)+\mathcal{O}(\al^2),\\
			&J_{+}^{(1)}(\al,0)= \dfrac{\pi^2}{6}- \dfrac{\al^2}{2}\left(\ln\dfrac{\al}{\pi}+C-\dfrac{1}{2}\right)+\mathcal{O}(\al^2).\\
		\end{cases}
	\end{align}
	
This process has the contribution from five Higgs particles in total into the effective potential. However, there are only two scalar Higgs particles $h, H$, which are associated with the two nonzero vacuum expectation values $v_1, v_2$.  Therefore, the minimum conditions then are
	
	\begin{align}
		&V_{eff}(\mathcal{V}_0,0)=0,\qquad
		\dfrac{\pa V_{eff}(\mathcal{V},0)}{\pa \mathcal{V}}\bigg|_ {\mathcal{V}=\mathcal{V}_0}=0,\\
		&\dfrac{\pa^2 V_{eff}(\mathcal{V},0)}{\pa \mathcal{V}^2}\bigg|_ {\mathcal{V}=\mathcal{V}_0}=\left[m_{h}^2(\mathcal{V})+  m_{H}^2(\mathcal{V})\right]_{\mathcal{V}=\mathcal{V}_0}.
	\end{align}

With the minimum conditions, expanding the functions of  $J_{\pm}$, the effective potential can be rewritten as
	
	\begin{align}\label{bl}
		V_{eff}(\mathcal{V})= \dfrac{\la_T}{4}\mathcal{V}^4 -\theta T\mathcal{V}^3 + \ga( T^2 - T_0^2) \mathcal{V}^2,
	\end{align}
where,
	\begin{align}
		\fontsize{11pt}{15pt}\selectfont
		\begin{split}
			&\la_T=\dfrac{m_{h}^2(\mathcal{V}_0)+  m_{H}^2(\mathcal{V}_0) }{2\mathcal{V}_0^2} \bigg\{ 1+ \dfrac{1}{8\pi^2 \mathcal{V}_0^2 \left[m_{h}^2(\mathcal{V}_0)+  m_{H}^2(\mathcal{V}_0)\right] } \times \\
			& \quad\bigg[
			6m_{W^\pm}^4(\mathcal{V}_0) \ln \dfrac{bT^2}{m_{W^\pm}^2(\mathcal{V}_0)} +3m_{Z}^4(\mathcal{V}_0) \ln \dfrac{bT^2}{m_{Z}^2(\mathcal{V}_0)} + 2m_{H^\pm}^4(\mathcal{V}_0) \ln \dfrac{bT^2}{m_{H^\pm}^2(\mathcal{V}_0)} \\
			& \quad+ m_{h}^4(\mathcal{V}_0) \ln \dfrac{bT^2}{m_{h}^2(\mathcal{V}_0)} + m_{H}^4(\mathcal{V}_0) \ln \dfrac{bT^2}{m_{H}^2(\mathcal{V}_0)} + m_{A}^4(\mathcal{V}_0) \ln \dfrac{bT^2}{m_{A}^2(\mathcal{V}_0)} -12m_{t}^4(\mathcal{V}_0) \ln \dfrac{b_FT^2}{m_{t}^2(\mathcal{V}_0)} \bigg]\bigg\},\\
			&\theta= \dfrac{1}{12\pi \mathcal{V}_0^3}\bigg[ 6m_{W^{\pm}}^3(\mathcal{V}_0) +3m_{Z}^3(\mathcal{V}_0) +2m_{H^\pm}^3(\mathcal{V}_0)+ m_{h}^3(\mathcal{V}_0)+ m_{H}^3(\mathcal{V}_0) + m_{A}^3(\mathcal{V}_0) \bigg] ,\\
			& \ga = \dfrac{1}{24 \mathcal{V}_0^2 }\bigg[ 6m_{W^{\pm}}^2(\mathcal{V}_0) +3m_{Z}^2(\mathcal{V}_0) +2m_{H^\pm}^2(\mathcal{V}_0)+ m_{h}^2(\mathcal{V}_0)+ m_{H}^2(\mathcal{V}_0) + m_{A}^2(\mathcal{V}_0) +6 m_t^2(\mathcal{V}_0)  \bigg],\\
			& T_0^2 = \dfrac{1}{4\ga}\bigg\{m_{h}^2(\mathcal{V}_0)+  m_{H}^2(\mathcal{V}_0)  -\dfrac{1}{8\pi^2 \mathcal{V}_0^2}\bigg[ 6m_{W^{\pm}}^4(\mathcal{V}_0) +3m_{Z}^4(\mathcal{V}_0) +2m_{H^\pm}^4(\mathcal{V}_0)\\
			&\qquad\qquad + m_{h}^4(\mathcal{V}_0)+ m_{H}^4(\mathcal{V}_0) + m_{A}^4(\mathcal{V}_0) -12 m_t^4(\mathcal{V}_0)  \bigg]\bigg\}.
		\end{split} \label{44}
	\end{align}
	
The critical temperature $T_c$ is given by
	\be T_c=\fr{T_0}{\sqrt{1-\theta^2/[\ga\la_{T_c}}]},\label{th}
	\ee
	and the critical VEV can be derived as
	\be\mathcal{V}_c=\fr{2\theta T_c}{\la_{T_c}}.\ee	
	Therefore, the strength of EWPT is
	\be
	S=\fr{\mathcal{V}_c}{T_c}=\fr{2\theta}{\la_{T_c}}.
	\ee
	
Next, taking into account daisy loops,  the effective potential will have the form:
		\begin{align}
			V^{daisy}_{eff}=V_{eff}(\mathcal{V})-V^d(\mathcal{V}),\label{daisyloop}
		\end{align}
in which the second component on the right-hand side of Eq.~(\ref{daisyloop}) is the contribution of daisy loops \cite{carrington,curtin,katz} (especially the appendix A in Ref.~\cite{katz}). Here,  degrees of freedom are given by: $g_Z=3, g_W=6, g_h=g_A=g_H=1, g_{H^{\pm}}=2$ and
		\begin{align}
			V^d(\mathcal{V})=\fr{T}{12\pi}\sum_{i=h,W,Z,A,H,H^\pm}g_i \left \{\left[\fr{m_i^2(\mathcal{V}_0)
				\mathcal{V}^2}{\mathcal{V}^2_0}+\Pi_i(T)\right]^{3/2}-\fr{m_i^3(\mathcal{V}_0) \mathcal{V}^3}{\mathcal{V}^3_0}\right \},
		\end{align}
		\begin{align}
			&\Pi_W(T)=\fr{22}{3}\fr{m_W^2(\mathcal{V}_0)}{\mathcal{V}^2_0}T^2, \nonumber\\
			&\Pi_Z(T)=\fr{22}{3}\fr{(m_Z^2(\mathcal{V}_0)-m_W^2(\mathcal{V}_0))}{\mathcal{V}^2_0}T^2,\nonumber\\
			&\Pi_h(T)=\fr{2m_W^2(\mathcal{V}_0)+m_Z^2(\mathcal{V}_0)+m_h^2(\mathcal{V}_0)+2m_t^2(\mathcal{V}_0)}{4\mathcal{V}^2_0}T^2+(\La_{hH}+\La_{hA}+\La_{hH^\pm}).T^2. \label{2.164}
		\end{align}
As for exotic Higgs particles then
		\begin{align}
			&\Pi_{h-exotic}= (\La_{hH}+\La_{hA}+\La_{hH^\pm}).T^2\\
			&\Pi_A(T)\sim \fr{m^2_A(\mathcal{V}_0)}{\mathcal{V}_0^2}T^2, \nonumber\\
			&\Pi_H(T)\sim \fr{m^2_H(\mathcal{V}_0)}{\mathcal{V}_0^2}T^2,\nonumber\\
			&\Pi_{H^{\pm}}(T)\sim \fr{m^2_{H^{\pm}}(\mathcal{V}_0)}{\mathcal{V}_0^2}T^2. \label{2.164b}
	\end{align}
	
$\La_{hH}+\La_{hA}+\La_{hH^\pm}$ are coefficients representing the contribution from exotic Higgs daisy loops to SM-like Higgs boson. $A$, $H$, $H^\pm$ are called exotic Higgs for short.
	
The daisy loops of exotic Higgs boson can be omitted, since these masses of particles are large and $m(\mathcal{V})/T\sim 1$ \cite{carrington}. This can be explained in Sec.~\ref{VF}.
	
Note that $\mathcal{V}$ can be  $v_1, v_2$ or $v$. Let  $v^2=a.v^2_2$, the relation between $\tan\beta$ and $a$ is
	\be
		\tan\beta=\sqrt{1/(a-1)}.
	\ee
	
From this, the masses of the particles in terms of $a$ are given in Table \ref{mass12}.
	
	\begin{table}[htp]
		\centering
		\begin{tabular}{ c|c|c|c }
			\hline
			Particles & $m^2(v_1,v_2)$ & $m^2(v_2)$ & $m^2(v_1)$  \\
			\hline
			$m^2_{W^\pm}$ & $\fr{g^2v^2}{4}$ &$m^2_{W^\pm}/a$ &$m^2_{W^\pm}(v_2).(a-1)$\\
			
			$m^2_Z$ & $(g^2+ g'^2)\fr{v^2}{4}$ & $m^2_Z/a$ & $m^2_{Z}(v_2).(a-1)$  \\
			
			$m^2_h$ & $ \fr{1}{4} f_h v^2$ & $m^2_h/a$ & $m^2_{h}(v_2).(a-1)$  \\
			
			$m^2_H$ & $ \fr{1}{4} f_H v^2$ & $m^2_H/a$ & $m^2_{H}(v_2).(a-1)$ \\
			
			$m^2_A$ & $ - \dfrac{1}{2} (2l_2 - l_3) v^2  $ & $m^2_A/a$ & $m^2_{A}(v_2).(a-1)$   \\
			
			$m^2_{H^\pm}$ & $ - l_2 v^2$  & $m^2_{H^\pm}/a$ & $m^2_{H^{\pm}}(v_2).(a-1)$  \\
			
			$m_t^2$ & $f_t^2 v^2$ & $m^2_t/a$ & $m^2_{t}(v_2).(a-1)$ \\
			\hline
		\end{tabular}
		\caption{Squared mass of the gauge bosons and scalar bosons in 2HDM-$S_3$.}
		\label{mass12}
	\end{table}
	
With $v_1$ different from $v_2$, this model has two stages of phase transition. We assume that $v_1<v_2$, which means $1<a<2$ or $1<\tan\beta$. As the above sections pointed out, in 2HDM-$S_3$, the particles’ masses can be changed such that there are no mixing terms between $v_1$ and $v_2$. So the correct effective potential for this model is
	\be
		V_{eff}^{S_3}=V_{eff}^{S_3}(v_1)+V_{eff}^{S_3}(v_2).
	\ee
	
With the above formula for the potential, the phase transition’s strength does not depend on $a$.

\subsection{Probing the independence of EWPT strength on $\tan\beta$}
	
Note that the EWPT in this model occurs in two stages and the mass components of the involved particles are given in Table \ref{mass12}. Let us assume that the phase transition’s strength of the first stage has already been calculated, $S_1>1$ for $\mathcal{V} = v_2$.	
	
It would like to prove that the second phase transition’s strength ($S_2$), corresponds to  $\mathcal{V}^2 = v_1^2 = v_2^2 (a-1)$ will actually not change, that it is still equal to $S_1$. Or in other words, it does not depend on $a$.
	
To do this, the functions $\la_{T_c}(v_1), \theta(v_1), \ga(v_1), T_c(v_1)$ must be indicated the independent of $a$ (equal to themselves when calculated with $v_2$). First, let us consider the function $\theta$ correspond to $\mathcal{V} = v_1$,
	\be
		\theta (v_1)= \dfrac{1}{12\pi v_1^3}\bigg[ 6m_{W^{\pm}}^3(v_1) +3m_{Z}^3(v_1) +2m_{H^\pm}^3(v_1)+ m_{h}^3(v_1)+ m_{H}^3(v_1) + m_{A}^3(v_1) \bigg].
	\ee
The masses in the bracket and $v_1$ all have the same power of 3. Hence they are all proportional to $(a-1)^{3/2}$. By extracting this factor out of the masses and canceling it with the exact same factor from  $v_1$ in the denominator, $\theta(v_1)$ is independent of $a$, or $\theta (v_1) = \theta (v_2)$. Similarly $\ga(v_1)$ does not depend on $a$, but $T^2_0 (v_1)$ depends on $(a-1)$, specifically $T^2_0 (v_1) = (a-1) T^2_0 (v_2)$. Because of this dependence on $(a-1)$ of $T_0^2 (v_1)$, $\la_{T_c}(v_1)$ will not depend on $a$. The proof is as follows. Consider the function $\la_{T_c}$ corresponds to $v_1$,
	\begin{align}
		\fontsize{11pt}{15pt}\selectfont
		\begin{split}
			\la_{T_c} (v_1) &=\dfrac{m_{h}^2(v_1)+  m_{H}^2(v_1)}{2v_1^2} \bigg\{ 1+ \dfrac{1}{8\pi^2 v_1^2 \left[m_{h}^2(v_1)+  m_{H}^2(v_1)  \right] } \times \\
			& \quad\bigg[
			6m_{W^\pm}^4(v_1) \ln \dfrac{b{T_c}^2(v_1)}{m_{W^\pm}^2(v_1)} +3m_{Z}^4(v_1) \ln \dfrac{b{T_c}^2(v_1)}{m_{Z}^2(v_1)} + 2m_{H^\pm}^4(v_1) \ln \dfrac{b{T_c}^2(v_1)}{m_{H^\pm}^2(v_1)} \\
			& \quad+ m_{h}^4(v_1) \ln \dfrac{b{T_c}^2(v_1)}{m_{h}^2(v_1)} + m_{H}^4(v_1) \ln \dfrac{b{T_c}^2(v_1)}{m_{H}^2(v_1)} + m_{A}^4(v_1) \ln \dfrac{b{T_c}^2(v_1)}{m_{A}^2(v_1)} \\ &-12m_{t}^4(v_1) \ln \dfrac{b_F{T_c}^2(v_1)}{m_{t}^2(v_1)} \bigg]\bigg\}.
		\end{split} \label{49}
	\end{align}
With $T_c$ is given by Eq.~(\ref{th}) in the no daisy loop case. By the same reasoning from above, $\la_{T_c} (v_1)$ will not depend on $a$ when the logarithmic factors do not depend on $a$. Indeed, we consider the general expression inside the logarithmic functions:
	\begin{align}
		\begin{split}
			\fr{T_c^2 (v_1)}{m^2(v_1)} &= \fr{T_0^2 (v_1)}{\left[  1 - \theta^2(v_1)/\ga(v_1) \la_{T_c}(v_1)\right] m^2(v_2)}\\
			&= \fr{(a-1) T^2_0 (v_2)}{\left( 1 - \theta^2(v_2)/\ga(v_2) \la_{T_c}(v_1)\right) (a-1)m^2(v_2)}\\
			&= \fr{T^2_0(v_2)}{\left( 1 - \theta^2(v_2)/\ga(v_2) \la_{T_c}(v_1)\right) m^2(v_2)},
		\end{split} \label{50}
	\end{align}
in which the functions $\theta(v_1), \ga(v_1)$ are all independent of $a$ as proven earlier. The expression inside the logarithmic functions actually depends on $\la_{T_c}(v_1)$. Substitute the expression Eq.~(\ref{50}) into Eq.~(\ref{49}), we can finally realize that $a$ no longer appears in the expression Eq.~(\ref{49}). This proves that $\la_{T_c}(v_1)$ does not depend on $a$. So, the functions $\la_{T_c}(v_1), \theta(v_1), \ga(v_1), T_c(v_1)$ is truly independent of $a$ when the effective potential without daisy loops.
	
When the effective potential with daisy loops which is Eq.~\eqref{daisyloop}, the critical temperature $T_C(v_1)$ is not Eq.~\eqref{th} but $T_C(v_1)\sim T_0(v_1)$. So $\la_{T_C}(v_1)$ still does not depend on $a$. Furthermore, by the similar proof, $V^d(v_1)$ does not depend on $a$. Because $V^d(v_1)$ just depends on the ratio $m^2(\mathcal{V})/\mathcal{V}^2$. So finally at the critical temperature, the effective potential with daisy loops $[V^{daisy-S_3}_{eff}(v_1)]$ remains independent of $a$ and deduced that $S$ in the same regardless of being calculated with $v_1$ or $v_2$.
	
This result agrees with the conclusions in Ref.~\cite{davidson}, which concludes that $\tan\beta$ is not a meaningful parameter in the 2HDM.

Also commented in Sec.~\ref{IIE}, in the Fig. 1 of Ref.~\cite{dori}, in the 2HDM model, since the ratio between the two phase transition strengths may not depend much on $\tan\beta$ . In other words, the strength of the phase transition can be independent of $\tan\beta$. We have clearly demonstrated this in the 2HDM-$S_3$ model, thanks to the $S_3$ symmetry that separates the two phase transitions.
	
\subsection{The true critical temperatures}
	
To indicate critical temperatures in the model, the effective potential without daisy loops is only used. The estimation of daisy loop contributions will be done in Sec.~\ref{VF}. As analyzed in Sec.~\ref{stewpt}, the 2HDM-$S_3$ model will have two critical temperatures which correspond to the two stages.
	
Since the coefficients are independent of $a$ as shown above, the parameters of the second phase transition can be expressed in terms of the parameters of the first phase transition,
	\begin{align}
		\la_{T,v_1}&=\la_{T,v_2}=\la_{T,v},\\
		\theta_{v_1}&=\theta_{v_2}=\theta_{v},\\
		\ga_{v_1}&=\ga_{v_2}=\ga_{v},.\\
		T^2_{0,v_2}&=T^2_{0,v_1}/(a-1)=T^2_{0,v}/a.
	\end{align}
With these equalities, the effective potentials of the second stage and the combined stages can be expressed in terms of the effective potential of the first phase transition stage
	\begin{align}
		V_{eff}^{S_3}(v_2)&= \dfrac{\la_T}{4}v_2^4 -\theta T_{v_2}v^3_1 + \ga(T^2_{v_1} - T_{0,v_1}^2)v_1^2,\\
		V_{eff}^{S_3}(v_1)&=\dfrac{\la_T}{4}v_1^4 -\theta T_{v_1}v^3_1 + \ga(T^2_{v_1} - T_{0,v_1}^2)v_1^2\\
		&=(a-1)^2\dfrac{\la_T}{4}v_2^4 -(a-1)^2\theta T_{v_2}v^3_2 + (a-1)^2\ga(T^2_{v_2} - T_{0,v_2}^2)v_2^2\\
		&=(a-1)^2V_{eff}(v_2),\\
		V_{eff}(v)&=\dfrac{\la_T}{4}v^4 -\theta T_{v}v^3 + \ga(T^2_{v} - T_{0,v}^2)v^2\\
		&=a^2\dfrac{\la_T}{4}v_2^4 -a^2\theta T_{v_2}v^3_2 + a^2\ga(T^2_{v_2}-T_{0,v_2}^2)v_2^2=a^2V_{eff}(v_2).
	\end{align}
Hence, $V_{eff}(v)\ne V_{eff}^{S_3}(v_1)+V_{eff}^{S_3}(v_2)=V^{S_3}_{eff}$. From these equations, $V_{eff}(v)$ can be deduced
	
	\be
		V_{eff}(v)=\fr{a^2}{(a-1)^2+1}V^{S_3}_{eff}=f(a)V^{S_3}_{eff}.
	\ee
Here when writing down the effective potential of the system in terms of $V_{eff}(v)$, the correct effective potential of the system has been multiplied by $\fr{a^2}{(a-1)^2+1}$ times.
	
	\begin{figure}[H]
		\centering
		\includegraphics[width=0.6\textwidth]{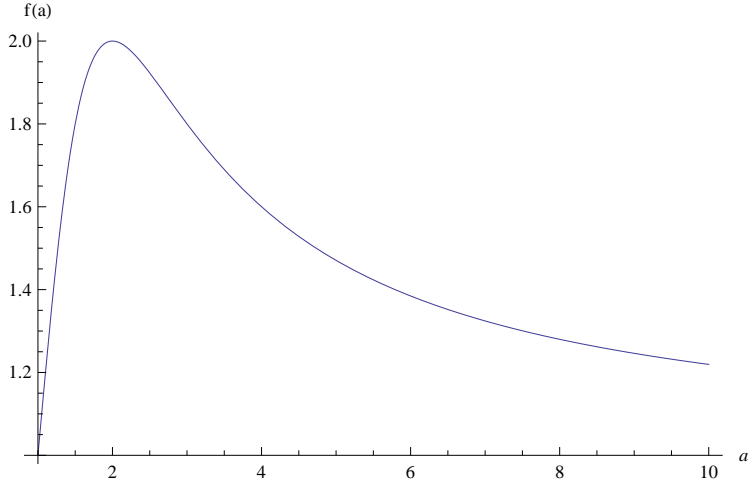}
		\caption{The function $f(a)$, the ratio 
		between $V^{S_3}_{eff}$ and $V_{eff}(v)$}
		\label{fig5}
	\end{figure}
	
According to Fig.~\ref{fig5}, it follows  $1<a<2$, and function $f(a)$  has value of $1$ at $a=1$, that is  $V_{eff}(v)=V^{S_3}_{eff}$, or $v_1=0$, but this cannot be true. The maximum value of function $f(a)$ is $2$ when $a=2$, which is $v_1=v_2$. Therefore when the two VEVs are equal to each other, the difference between the two effective potentials is at maximum. When $a>2$ or  $v_1>v_2$, it can exchange $v_1\longleftrightarrow v_2$, so that this case is similar to the case of $a<2$.
	
When the effective potential is rewritten as  $V_{eff}(v)$ when calculating the phase transition’s strength, our $S$ is correct. However, now the temperature of the phase transition $T_C$ of the system turns out not to be correct. If $v_1\ne v_2$, our system has two phase transition temperatures $T_{C1}<T_C$ and $T_{C2}<T_C$, respectively. Hence, writing down the effective potential for the system as $V_{eff}(v)$ is just a way to compare it to that of SM (or putting this model in the context of SM, we call it “the SM-like effective potential”), and now $T_C$ is not the true temperature for the phase transition of the system, and it should be called “the SM-like critical temperature.”

As commented earlier in Sec.~\ref{IIE}, when studying phase transitions in 2HDM, studies rarely mention phase transition temperature. Because the analysis of the phase transition temperature of the two VEVs model would be very difficult due to their mixing. As analyzed in this section, the $S_3$ symmetry separates the two phase transitions, so it makes the determination of the transition temperature more obvious.
	
	\begin{figure}[h!]
		\centering
		\begin{subfigure}{0.5\textwidth}
			\centering
			\includegraphics[width=1\linewidth]{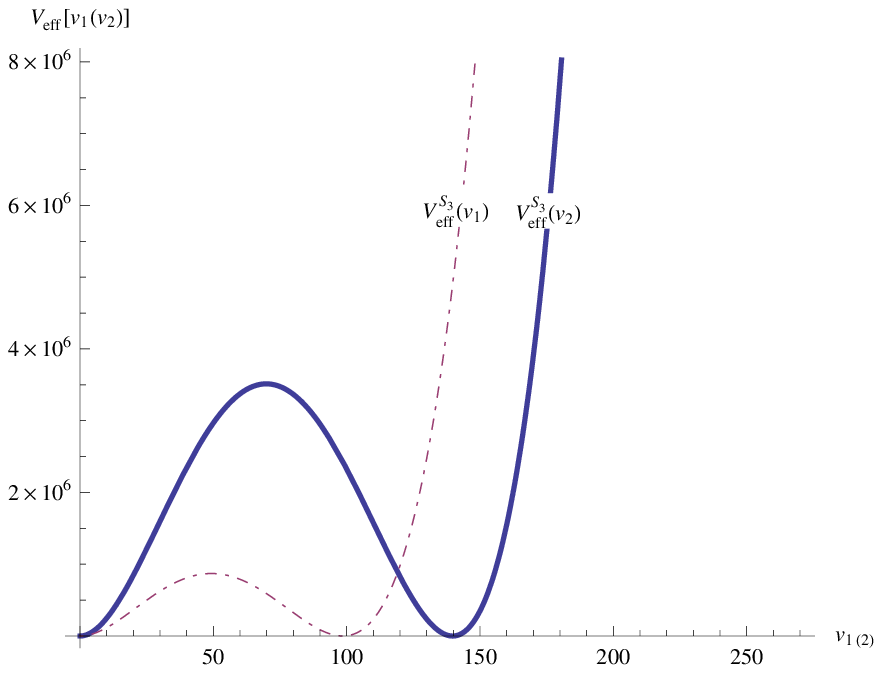}
			\caption{$V_{eff}^{S_3}(v_2)$ and $V_{eff}^{S_3}(v_1)$ are the first and second EWWPT when $a=3/2$ and $S=1$.}
			\label{fig6a}
		\end{subfigure}%
		\begin{subfigure}{0.5\textwidth}
			\centering
			\includegraphics[width=1\linewidth]{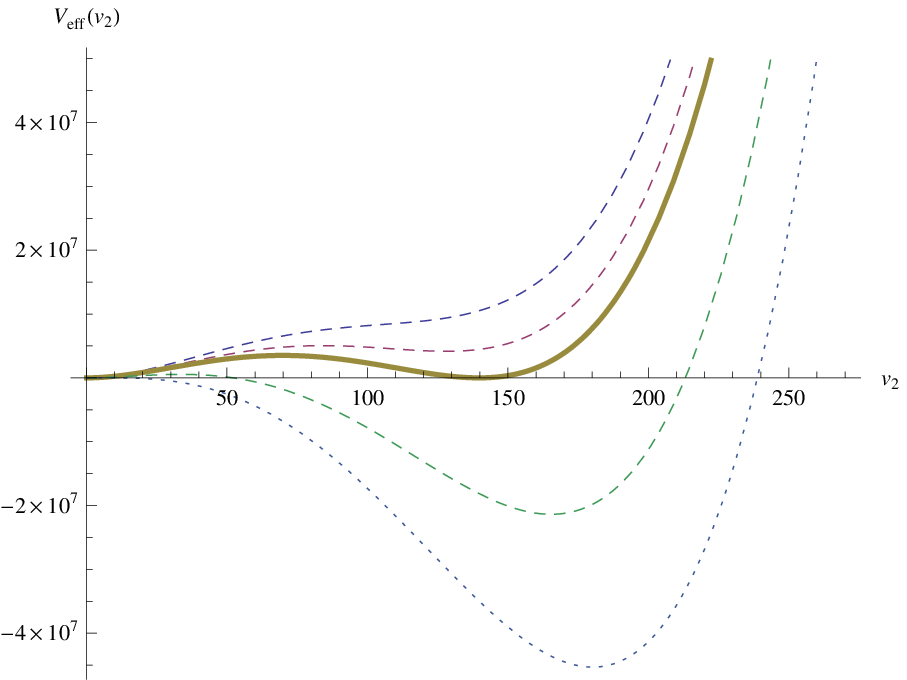}
			\caption{$V_{eff}^{S_3}(v_2)$ at several temperatures when $a=3/2$ and $S=1$.}
			\label{fig6b}
		\end{subfigure}
		\caption{The effective potential with $a=3/2$.}
		\label{fig6}
	\end{figure}
	
Fig.~\ref{fig6b} shows us the potential $V_{eff}^{S_3}(v_2)$ at different temperatures. These potentials all have the second non zero minima, with $m_{H}(v_2)=150~\textrm{GeV}$, $m_{H^{\pm}}(v_2)=m_{A}(v_2)= 302.087~\textrm{GeV}$. The solid line that corresponds to  $T_{C1}=139.739~\textrm{GeV}$, shows that there exists a potential well between the two minima. This is proof of the existence of the first-order in the phase transitions. 	
	
In Fig.~\ref{fig6a}, the solid line is the shape of $V_{eff}^{S_3}(v_2)$ at $m_{H}(v_2)=150~\textrm{GeV}$, $m_{H^{\pm}}(v_2)=m_{A}(v_2)=302.087~\textrm{GeV}$, and $T_{C1}=139.739~\textrm{GeV}$. The dash-dotted line is $V_{eff}^{S_3}(v_1)$ when $m_{H}(v_1)=150/\sqrt{2}~\textrm{GeV}$, $m_{H^{\pm}}(v_1)=m_{A}(v_1)=302.087/\sqrt{2}~\textrm{GeV}$, and $T_{C2}=98.761~\textrm{GeV}$. The nonzero minimum and maximum of the dash-dotted are smaller than that of the solid line and $T_{C1}>T_{C2}$, which shows that the phase transition must occur in two stages. The distance between the two stages is $\De T_C=T_{C1}-T_{C2}=40.978~\textrm{GeV}$.
	
\begin{figure}[H]
		\centering
		\includegraphics[width=0.5\textwidth]{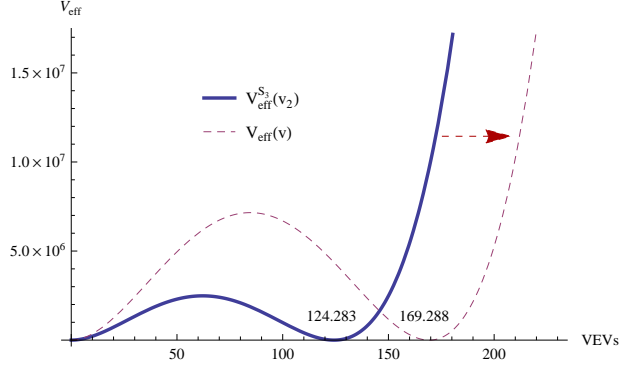}
		\caption{$V_{eff}^{S_3}(v_2)$ and $V_{eff}(v)$ when $v_1=v_2$ or $a=2, \tan\beta=1$}
		\label{fig7}
	\end{figure}

To be more intuitively in the comments, we plot the effective potential $V^{S_3}_{eff}(v_2)$ and $V_{eff}(v)$ in case of $a=2$ as in Fig.~\ref{fig7}. The blue solid line is the potential $V_{eff}(v_2)$ when $m_{H}(v_2)=140~\textrm{GeV}$, $m_{H^{\pm}}(v_2)=m_{A}(v_2)=272.647~\textrm{GeV}$ and the phase transition’s temperature $T_{C1}=124.283~\textrm{GeV}V$. The dashed line is for $m_{A}(v)=198~\textrm{GeV}$, $m_{H^{\pm}}(v)=m_{A}(v)=362.284~\textrm{GeV}$ and the critical temperature is $T_C=169.288~\textrm{GeV}$.
	
Through Fig.~\ref{fig7}, when $a=2$, that is when $v_1=v_2$, the two stages of the EWPT occur at the same time. Each stage of the phase transition is described by the blue solid line. The correct temperature for the phase transition of the system must be $T_{C1}$.
	
However, also through Fig.~\ref{fig7}, if we describe our system using the composite potential $V_{eff}(v)$, that is if we study the phase transition of the system in just one stage as in SM, the effective potential of the system at the phase transition temperature is as the dashed line. Here the temperatures for the system are $T_C>T_{C1}$. By describing the EWPT in only one stage as in the SM, the two effective potentials were turned in the solid line into the dashed line. Accidentally, this did not change the strength of the phase transition but instead increased the critical phase transition’s temperature and VEV, making them different compared to the correct ones.

\subsection{Searching the first-order EWPT and the role of $\tan\beta$}\label{VF}
	
In order to meet a first-order phase transition, the transition strength must have its value bigger or at least equal to 1. However, there are three unknown variables $m_{A}, m_{H^{\pm}}$ and $ m_{H}$ in our problem. Therefore we can assume that $m_{A}\equiv m_{H^\pm}$. This assumption is only intended to reduce the number of variables in the problem and find the domain for the masses of the particles in the first-order phase transition. This assumption should not be applied to the parameters in the Higgs potential. Also assuming that $m_{A}=m_{H^\pm}$ or $m_{H}=m_{H^{\pm}}$, but the results are all the same. Choosing to use $m_{A}= m_{H^\pm}$ which is consistent with the previous studies and the data for the parameter $\rho$ \cite{michela1,michelb1,michelb2,michelb3,michelb4}. 

According to the comments in Sec.~\ref{IIE}, especially in Ref.~\cite{dori}, there can be many suggestions between the three quantities $m_{H^\pm}, m_{H}, m_{ A}$ but it is possible for a first order EWPT, so from the suggestions, we can choose the scenario $m_{A}= m_{H^\pm}$ in the 2HDM-$S_3$ model. Because the symmetry $S_3$ does not lose or add any of the three additional Higgs bosons.

Furthermore, the effective potential without daisy loop is first used to calculate $S$. Then daisy loops will be additionally calculated later.
	
\begin{figure}[H]
		\centering
		\includegraphics[width=0.7\textwidth]{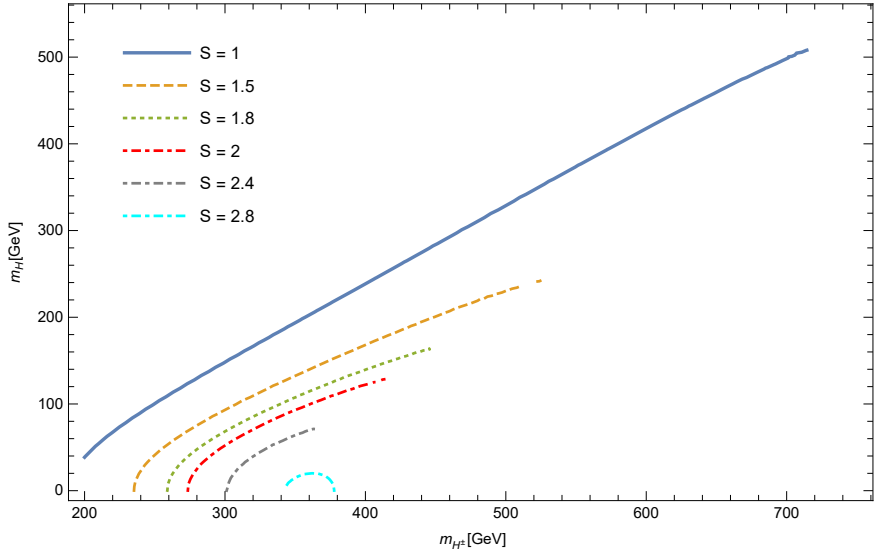}
		\caption{The strength of first stage EWPT with $a=3/2$}
		\label{fig8}
	\end{figure}

From there, the domain for the value of the masses cannot be too broad, in fact, these domains must be closed. Indeed, to have $S>1$, and if we apply the following conditions altogether: $T^2_0>0$ and $T_C$ must be real or according to Eq.~(\ref{th}), $1-\theta^2/[\ga\la_{T_c}]>0$, mass domains must be closed. The numerical solution for the case of $a=3/2$ for the first stage of the phase transition that corresponds to $v_2$ is given as Fig. \ref{fig8}.
	
The contours of $S=1$ are plotted onto the mass axes and then gradually increase the value of $S$. With different values of $a$, a range of values of the phase transition strength was find, $1\leq S <2.8$.
	
\begin{figure}[H]
		\centering
		\includegraphics[width=0.7\textwidth]{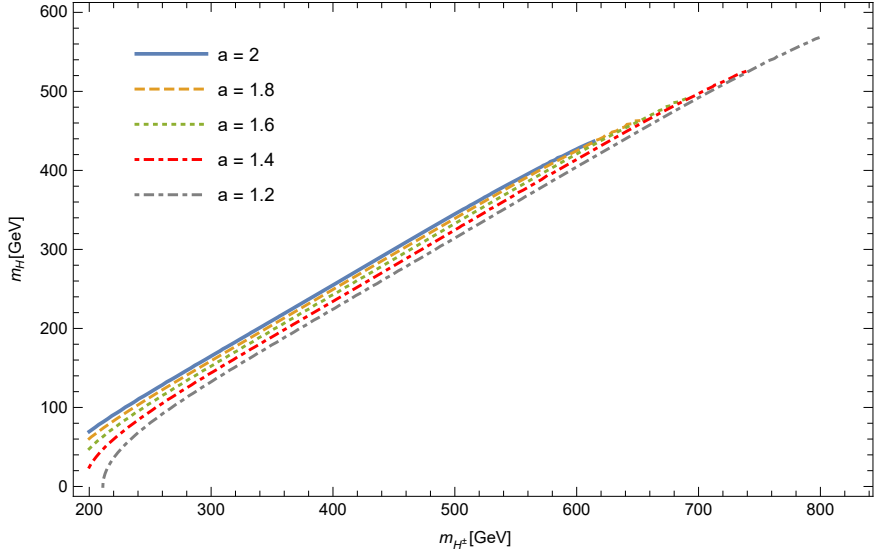}
		\caption{The mass domain of the particles depends on $a$ with $S=1$ in the first phase transition}
		\label{fig9}
	\end{figure}

Moreover, according to Fig.~\ref{fig8}, when $S$ increases, the domain for the masses of the particles must narrow down. Therefore, in order to find these domains for different values of $a$, only plotting $S=1$ with different $a$’s, which is shown in Fig.~\ref{fig9}.
	
	Carefully looking at Fig.~\ref{fig9}, when $a$ increases, the mass domains of the particles narrow down. Since $1<a\le 2$, the mass domain for each value of $a$ as follows:
	\begin{align}
		200 \quad \textrm{GeV}  <& m_{H^\pm}(v_2)<800 \quad \textrm{GeV},\label{k09}\\
		0<&m_{H}(v_2)<600 \quad \textrm{GeV}.\label{k10}
	\end{align}

The second stage of the phase transition that corresponds to $v_1$ is similar to the first stage. And because the strength of the phase transition does not depend on $a$, if the first stage of the phase transition has the transition’s strength larger than 1 then so does stage 2. Therefore, the domains for the masses of the particles at stage 2 can be indicated,
	
	\begin{align}
		200\times \sqrt{(a-1)} \quad \textrm{GeV} <& m_{H^\pm}(v_1)<800\times \sqrt{(a-1)} \quad \textrm{GeV},\label{k11}\\
		0<&m_{H}(v_1)<600\times \sqrt{(a-1)} \quad\textrm{GeV}.\label{kl2}
	\end{align}

Thus, according to Eq.~\eqref{k10} and Eq.~\eqref{kl2}, although $a$ (or $\tan\beta$) does not affect the phase transition’s strength, it affects the domains of the masses of the particles to have the first-order phase transition.
	
Combining Eqs.~\eqref{k09}, \eqref{k10}, \eqref{k11}, \eqref{kl2} together, it follows that
	\begin{align}
		200\times \sqrt{a} \quad \textrm{GeV} < &m_{H^\pm}<800\times \sqrt{a} \quad \textrm{GeV},\label{k11s}\\
		0<&m_{H}<600\times \sqrt{a} \quad \textrm{GeV}.\label{kl2s}
	\end{align}
Notice that the maximum value of $a$ is 2, it follows from Eqs.~\eqref{k11s}, \eqref{kl2s} that for $a=\sqrt{2}$. So, the maximum value of masses are only increased by about $1.41$ times. Therefore, the effect of $a$ on the mass domain of the particles is not too large. This is also easy to see when observing the lines in Fig.~\ref{fig9}, they are very close together.	
	
One more thing, if $a$ is closer to 2, the two stages are also closer to each other. Hence, $a$ can be used to define the distance between the two-phase transition stages in this model.

With the analysis of EWPT in 2HDM, the value of $\tan\beta$ is quite wide (from 1 to 17 in all scenarios as indicated in Sec.~\ref{IIE}), it is almost a free parameter in the EWPT problem. However, in the 2HDM-$S_3$ model, due to the $S_3$ symmetry, the two EWPT stages are separated, thereby highlighting the role of $\tan\beta$ which determines the width of the mass domain as well as the gap between the two stages.
	
The last important part in this section will be estimating the contribution from daisy loops. Based on Eq.~\eqref{kl2}, the masses of the exotic Higgs bosons are usually chosen to be larger than the mass of the top quark. This is in keeping with the difficulty of detecting these particles today. Because it only consider the temperature region where EWPT occurs or $\fr{\mathcal{V}_c}{T_c}>1$. So $\fr{m_{A,H,H^\pm}(\mathcal{V}_c)}{T_c}\sim 1$, the contribution of daisy loops of exotic Higgs bosons will be small \cite{carrington}. Thus when adding daisy loops, only the daisy loops of gauge boson and SM-like Higgs boson, i.e., Eqs.~\eqref{2.164} are taken into account and neglecting $\La_{h-exotic}$. But notice that, in the temperature region $T\gg T_C$, the contributions from daisy loops of the exotic Higgs bosons cannot be ignored.
	
\begin{figure}[H]
		\centering
		\includegraphics[width=0.7\textwidth]{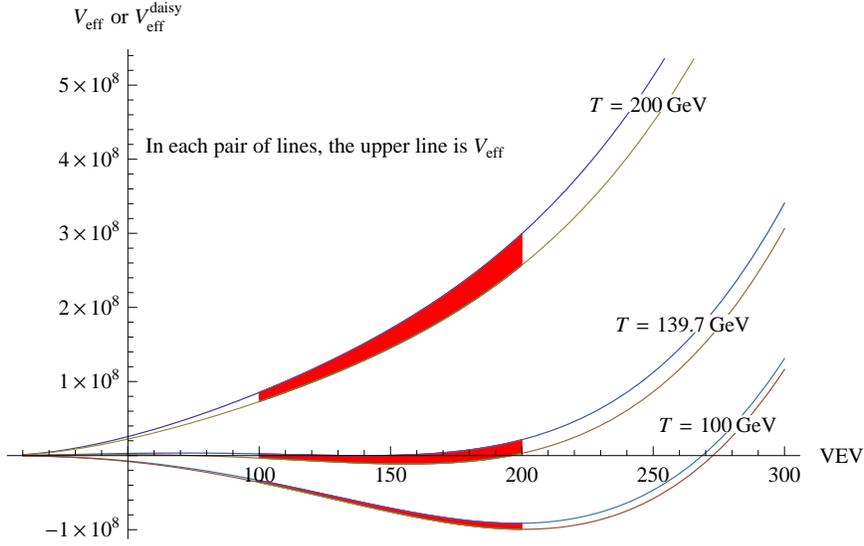}
		\caption{Difference between $V^{S_3}_{eff}$ and $V_{eff}^{daisy}$ with $m_{H}(v_2)=150~\textrm{GeV}, m_{H^{\pm}}(v_2)=m_{A}(v_2)=302.087~\textrm{GeV}.$}
		\label{fig10}
	\end{figure}

Take a look at the graphs in Fig.~\ref{fig10}, the red zone indicates the difference between $V_{eff}$ and $V_{eff}^{daisy}$, the higher the temperature, the larger the difference, and the larger the area of the red region. The phase transition temperatures in this model are similar to SM, they are in the range of $100$ to $150~\textrm{GeV}$. Because $v=\sqrt{v^2_1+v^2_2}=246~\textrm{GeV}$ and $v_1, v_2<246~\textrm{GeV}$. So also from Fig.~\ref{fig10}, when $T<150~\textrm{GeV}$, the area of the red region is very small.
	
Thus, also from Fig.~\ref{fig10}, daisy loops will increase the phase transition temperature. Indeed, the second pair of lines in Fig.~\ref{fig10}, when $T=139.739~\textrm{GeV}$ is also the phase transition temperature corresponding to $V_{eff}$ (the upper line). But $V_{eff}^{daisy}$ (bottom line) has a second minimum that is below the VEV axis, so $T=139.739~\textrm{GeV}$ is not yet the phase transition temperature corresponding to $V_{eff}^{daisy }$, this phase transition temperature must be greater than $139.739~\textrm{GeV}$. Finally, a sure result is that as the phase transition temperature increases, the phase transition strength will decrease.
	
Next to see the effect from daisy loops about $S$ as the above comments. The masses of $H^{\pm}$ and $H$ are randomly selected, then recalculate $S$ with and without daisy loops, as shown in Table \ref{hinhv}.
	\begin{table}[!ht]
		\centering
		\begin{tabular}{c|c|c|c|c}\hline
			$m_{H}(v_2)$ [GeV]&$m_{H^{\pm}}(v_2)$ [GeV]&$S_{nodaisy}$&$S_{daisy}$&$S_{daisy}/S_{nodaisy}$\\ \hline
			300 & 500 &  1.116 & 1.0938 & 0.979\\
			400 & 600 &  1.048 & 1.030 & 0.982\\
			250 & 470 &  1.253 & 1.224 & 0.976\\
			220 & 430 &  1.254 & 1.222 & 0.974\\
			180 & 410 &  1.452 & 1.410 & 0.971\\	
			\hline
		\end{tabular}
	\caption{Strength $S$ with and without daisy loops}\label{hinhv}
	\end{table}
	
The ring-loops reduce $S$ by $2/3$ times \cite{r23, 1101.4665}. But in this model, this ratio is about $0.97$ times or the maximum value of strength is about $2.71$. However, this still ensures the first-order EWPT.
	
Although the daisy loops do not significantly change the strength of the phase transition in this model, but the role of reducing divergences in effective potential calculations cannot be ignored.
	
In addition, it should be noted that in this model the triggers for the first order phase transitions are heavy Higgs particles. So the daisy loops of heavy Higgs have no effect in EWPT. However, other models exist light particles besides the SM, surely the daisy loops of these particles will have a great effect on EWPT.
	
\section{CONCLUSION AND OUTLOOKS}\label{vi}
	
The 2HDM-$S_3$ was chosen to analyze the EWPT stages, not only because it is close to the SM but also because the structure of the EWPT process in this model is interesting enough for us to find new things. Moreover, although these models have some features that are new to the SM, they are far more complicated as they have more new fields and VEVs.
	
The symmetry breaking process that corresponds to the EWPT in the 2HDM-$S_3$, when compared with the SM, is depicted in the below diagram.
	\begin{center}	
	\begin{align*}
		&\text{2HDM-$S_3$}&\\
		&\Downarrow \text{breaking $v_2$}\\
		&\Downarrow \text{breaking $v_1$} \\
		&U(1)_Q&
	\end{align*}
\end{center}
We first summarize the structural analysis methods of the EWPT from previous studies along with the method in this article as follows:
	
	\begin{itemize}
		\item The first method as in Refs.~\cite{thdm,Fuyutob}: The EWPT process is considered as in the SM. The effective potential $V_{eff}(v)$ is used and $v^2=v^2_1+v_2^2$.
		\item The second method as in Refs.~\cite{dori,mayumi}: Assuming the equation Eq.~\eqref{1loopsca}, examining the multistep
		EWPT process, and the effective potential is a function of two variables $v_ 1$ and $v_ 2$. But when calculating the strength of phase transition $\xi=\fr{v_c}{T_c}, v^2=v^2_1+v^2_2$, it is still referred to as one phase as the standard model.
		\item The third method is in this article: Analyzing the division of the effective potential into two separate parts $[V_{eff}=V_{eff}(v_1)+V_{eff}(v_2)]$ and examining each stage separately.
	\end{itemize}

The EWPT has been intensively  considered in the 2HDM. Some remarks are in order.

In Refs. \cite{124,127}, the EWPT has been considered in the 2HDM type I and II. In these papers, the masses of heavy particles are not larger than 1 TeV and their mass difference is not bigger 400 GeV. In our results, the mass region of heavy particles ranges from 200 to 800 GeV. Therefore, the maximum mass difference between the heavy particles is about 600 GeV. The lattice simulation with one-loop effective potential for crystal has been considered in Ref. \cite{127}, the first order EWPT happens for a scenario $m_{H^\pm}=m_A$,  and  this agrees with our assumption. Therefore, the results in Refs. \cite{124,127} and ours are compatible.

It was shown that  the EWPT is related to a  significant uplifting of the Higgs vacuum \cite{125}. The first order phase transition leads to  the following condition
\be m_A > m_H + m_Z\,.
\ee
The mass domain of particles in our calculation is also compatible with this.

The first order EWPT is possible if $ 580\, \textrm{GeV} < m_{H^\pm } < 1 \, \textrm{TeV}$ \cite{126}.  This constraint agrees with our result $m_ {H^\pm}  < 800 \times  \sqrt{a} \, \textrm{GeV} \sim 1 $ TeV as in Eqs. \eqref{k11s} and \eqref{kl2s}.

In Ref. \cite{128}, the first order EWPT happens in the 2HDM type I and II with the mass difference  between $H,H^\pm$ and $A$ in the range (100, 300) GeV. This coincides with the mass region in  Eqs. \eqref{k09} and \eqref{k10}.  

From Fig. 1 in Ref. \cite{129} it follows that to have the first  order phase transition, the maximal mass difference among $m_H$ and  $m_A$ is about 500 GeV. In our study this value is about 600 GeV. Hence, both results are consistent.

In Ref. \cite{130}, Fig. 3, the effective potential is plotted in the region of masses lower than 600 GeV and the value of $\tan\beta$ runs from 1 to 20. The lines of the effective potential for different $\tan\beta$ values are very close to each other. The result shows that $S$ is almost independent on $\tan\beta$, and this supports our conclusion. 
	
All three methods are acceptable approximations. The effectiveness of the third method in this article is clearly stated in Sec.~\ref{v}. Mathematical techniques for analyzing VEVs in the first and second method have been successfully used to analyze decay channels in multi-VEV models \cite{hue,alves}.
	
Exploring the EWPT process into several stages has been analyzed in many other models besides 2HDM, as shown in Refs.~\cite{ptl,borges}. Therefore, the results of this paper aim to emphasize the feature of the multiphase in 2HDM-$S_3$ model and investigate the factors that affect the division of stages as well as the influence of $\tan\beta$.
	
The strength of the phase transition does not depend on $a$ or $\tan\beta$, and to study the full structure of the two stages of the phase transition, the effective potential must not be written as $V_{eff}(v)$. The greater $\tan\beta$ is, the narrower the mass domain of the particles in the first order of the phase transition becomes. The $S_3$ symmetry has shown that there are two subsequential stages in the phase transition process, which has not been shown clearly in the 2HDM where there are many mixing terms of the VEVs in the mass domain of the particles. 
	
The $S_3$ symmetry could explain the mixing of quarks or this symmetry could be related to the Yukawa couplings that can affect EWPT processes \cite{bpal1, bpal2, bpal3}. For example in Ref.~\cite{braconi}, changing the Yukawa coupling constants results in a first order EWPT. Therefore, the $S_3$ symmetry associated with the quark mixing has an effect on EWPT processes that need to be further elucidated after these works.
	
By analyzing the effect of daisy loops, a way of assessing the contribution from daisy loops: first the EWPT was calculated by using the effective potential without daisy loops, to estimate the mass domain of particles; then based on that mass domain, estimating the ratio $m(v)/T$ to consider the contribution from daisy loops.
	
In this paper, in order to reduce the number of variables in our problem, $m_A=m_{H^\pm}$ has been assumed. As said earlier, this assumption was only made to find the mass domain of the particles, and must not be applied to the parameters in the Higgs potential, since the real values of these 2 particles can be different from each other, even though their domains of mass can be identical. However, from Eq.~\eqref{blkl} and this assumption, $l_3\approx 0$. This assumption was made by the authors in order to satisfy the data of the parameter $\rho$ in the 2HDM \cite{michela1,michelb1,michelb2,michelb3,michelb4} so that if the 2HDM-$S_3$ also satisfies the data of the parameter $\rho$, this can lead to $l_3$ being very small. This is one of the results of this paper that can lead to research on the parameter $\rho$ in the 2HDM-$S_3$.	
	
As stated in remark 4, the assessment of the impact the first stage of EWPT has on the second stage of EWPT is made through the mixing terms of the VEVs. However, the investigation of the effects of $\ka$ is still unclear. All these effects of $\ka$ have been renormalized under the minimum conditions of the later EWPT stage. Therefore, from remark 4, to assess the effects of $\ka$, we must rebuild the whole effective potential that contains the mixing terms of VEVs all over. Assessing this direction is a new incoming, and interesting job after this paper. 	

Notice the comments in the Sec. \ref{IIE}, we focus on scalar decays into heavy fermions ($A, H, H^{\pm}\longrightarrow tt, tb$), which are the most promising channels for demonstrating the first-order EWPT through confirmation of additional bosons \cite{shinya}. Also, there is another way to check, we can measure the gravitational waves generated by the EWPT process in future experiments by LISA \cite{shinya}.
	
Finally, through the comments on effective potentials, writing down the effective potentials in the form of  $V_{eff}(v)$ is imprecise, but still, it is concise and gives accurate predictions for the strength of the phase transition. However, there will be some small errors in calculating the corresponding sphalaron energy, but these errors would not be large. Since the contributions of the effective potential term in the sphaleron energy are quite small, about 5.5\% \cite{phong2022}, when writing down the effective potential of the system in the form of $V_{eff}(v)=\fr{a^2}{(a-1)^2+1}V^{S_3}_{eff}$, this will make the sphaleron energy deviate by about 5.5\%. From this, to make the calculation of the sphaleron become more accurate, we only need to replace $V_{eff}(v)$ by $\fr{(a-1)^2+1}{a^2}V_{eff}(v)$, and the methods are still the same as in Refs.~\cite{Fuyutob,sphagt1,sphagt2,Fuyutoc}. In addition, if the two stages of the phase transition in this model occur at the same time or very close to each other, the bubbles of each of the phase transition stages can collide with each other or collide with the bubbles from other stages, thereby causing some big gravitational waves. For that matter, the full estimation of the contributing terms, as well as the impact of the ratio $\fr{(a-1)^2+1}{a^2}$ on the sphaleron energy for the gravitational wave calculation, will be the extension of this paper.

The method of high-temperature dimensional reduction to the 2HDM to obtain three-dimensional effective theories that can be used for nonperturbative simulations \cite{131}. These results can be used to recalculate EWPT in 2HDM, and to check our results. This is part of the upcoming work.
	
\section*{ACKNOWLEDGMENTS}
V.Q.P. would like to thank Pham Quang Khanh for reading and editing the article in English and running my small code. H. N. L. is thankful to Van Lang University. 
\appendix

\section{Effective action of multiscalar field models}\label{multi}

In the $\phi^4$ theory and the single field case, calculating the effective potential from summing the diagrams is shown in Ref.~\cite{quiros}. Let us consider a toy model described by two neutral scalar  fields ($\phi_i, i=1,2$)
with an action
\be
S(\phi_1,\phi_2)=\int d^4 x{\cal L}(\phi_1(x),\phi_2(x)),
\label{actsca}
\ee
where
\be {\cal L}(\phi_1(x),\phi_2(x)) = \sum_{i=1}^2 \left(\pa_\mu \phi^*_i \pa^\mu \phi_i - m^2_{\phi_i}
\phi_i^* \phi_i \right) - \fr{\la_\phi}4 \left(\phi^*_1 \phi_1 + \phi^*_2 \phi_2\right)^2
\ee

In the path-integral representation, the generating functional is as the following:
\be
Z[J]=\langle 0_{\rm out}\mid 0_{\rm in} \rangle_J \equiv \int
\mathcal{D}\phi_1 \mathcal{D}\phi_2\exp\{i(S(\phi_1,\phi_2)+\phi_1 J_1+\phi_2 J_2)\},
\label{zfunct}
\end{equation}
and \be
Z[J] \equiv \exp\{iW[J]\},
\label{wfunct}
\end{equation}
in which
\be
\phi_i J_i\equiv \int d^4x \phi_1(x) J_1(x); \quad i=1,2.
\label{product}
\end{equation}

The effective action $\Ga[\overline{\phi}_1,\overline{\phi}_2]$ is the Legendre transform of Eq.~(\ref{wfunct})
\be
\Ga[\overline{\phi}_1,\overline{\phi}_2]=W[J]-\sum_{i=1}^2\int d^4 x \fr{\De
W[J]}{\De J_i(x)} J_i(x)
\label{effaction}
\end{equation}

where the VEVs of $\phi_i$ are

\be
\overline{\phi_i}(x)=\fr{\de W[J]}{\de J_i(x)}
\label{phibar}
\end{equation}

From Eq.~(\ref{effaction}) and Eq.~(\ref{phibar}), the generating functionals can be obtained

\be
\fr{\de \Ga[\overline{\phi}_1,\overline{\phi}_2]}{\de \overline{\phi_i}}=
\fr{\de W[J]}{\de J_i}\fr{\de J_i}{\de
\overline{\phi_i}}- J_i-\overline{\phi_i}\fr{\de J_i}{\de
\overline{\phi_i}}= -J_i.
\label{j}
\end{equation}

where using of the notation from Eq.~(\ref{product}).
Eq.~(\ref{j}) implies in particular that

\be
\left.\fr{\de \Ga[\overline{\phi}_1,\overline{\phi}_2]}{\de \overline{\phi_i}}
\right|_{J_i=0}=0.
\label{vacuum}
\end{equation}

$Z[J]$ can be expanded in a power series of $J$ and in terms of Green functions $G_{(n)}$ as

\be
Z[J]= \sum^{\infty}_{n=0} \fr{i^n}{n!} \int d^4x_1 \ldots
d^4x_n J(x_1)\ldots J(x_n) G_{(n)}(x_1, \ldots,x_n)
\label{green}
\end{equation}

and

\be
iW[J]= \sum^{\infty}_{n=0} \fr{i^n}{n!} \int d^4x_1 \ldots
d^4x_n J(x_1)\ldots J(x_n) G^{\ c}_{(n)}(x_1, \ldots,x_n).
\label{greenc}
\end{equation}

However, in the next step, expanding the effective action in terms of the one-particle irreducible Green functions $(\Ga^{(n)})$,

\be
\Ga[\overline{\phi}_1,\overline{\phi}_2]= \sum^{\infty}_{n=0} \fr{1}{n!} \int d^4x_1 \ldots
d^4x_n \overline{\phi_1}(x_1)\ldots \overline{\phi_1}(x_{n_1}) \overline{\phi_2}(x_{n_1+1})\ldots \overline{\phi_2}(x_{n_2})
\Ga^{(n)}(x_1, \ldots,x_n).
\label{1pi}
\end{equation}

\begin{figure}[h!]
\begin{tikzpicture}[baseline=(current  bounding  box.center)]
\begin{feynman}
\vertex (x);
\vertex[right=1.75cm of x] (y);
\vertex[above right=1.25cm of x] (z);
\vertex[below right=1.25cm of x] (k);
\vertex[above left=of x] (a){\(\phi_1\)};
\vertex[below left=of x] (b){\(\phi_1\)};
\vertex[above right=of y] (c){\(\phi_1\)};
\vertex[below right=of y] (d){\(\phi_1\)};
\vertex[above left=of z] (e){\(\phi_2\)};
\vertex[above right=of z] (f){\(\phi_2\)};
\vertex[below left=of k] (g){\(\phi_2\)};
\vertex[below right=of k] (h){\(\phi_2\)};

\diagram*{
(z) --[scalar, quarter right] (x),
(y) --[scalar, quarter right] (z),
(k) --[scalar, quarter right] (y),
(x) --[scalar, quarter right] (k),
(x) --[scalar] (a),
(x) --[scalar] (b),
(z) --[scalar] (e),
(z) --[scalar] (f),
(k) --[scalar] (g),
(k) --[scalar] (h),
(y) --[scalar] (c),
(y) --[scalar] (d),
};
\end{feynman}
\end{tikzpicture}
\caption{The 1-loop diagram with the $\phi_1$ and $\phi_2$ external lines}\label{fig11}
\end{figure}
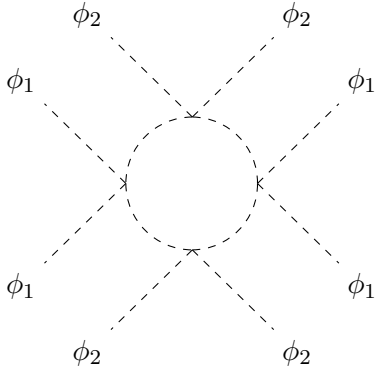

The number of vertices $n_1$ and $n_2$ are arbitrary but $n_1+n_2=2n$. We compute $\Ga^{(n)}(p_i=0)$ which are the diagrams with $2n$ external lines. Analyzing this in detail with the case of one-loop as shown Fig.~\ref{fig11} which is represented by the formula:

\begin{align}
\Ga_n(\phi_{1c},\phi_{2c})=&i\fr{1}{2n}\int \fr{d^4p}{(2\pi)^4}\left(\fr{i}{p^2-m^2(\phi_{1c},\phi_{2c})+i\varep}\right)^n(-i\la_{\phi_1})^{n_1/2}(\phi_{1c})^{n_1}(-i\la_{\phi_2})^{n_2/2}(\phi_{2c})^{n_2}.\label{dinh}
\end{align}

In Eq.~(\ref{dinh}), each vertex is a factor $-i\la_{\phi_i}$, the external line is the factor $\overline{\phi}_i=\phi_{ic}=const$. The above integral is easy to calculate if the external lines are of the same type (i.e., $n_1=0$ or $n_2=0$). However, when both $n_1$ and $n_2$ are nonzero, i.e., the external lines have both fields, calculating the above integral is not simple. Also then summing with $n=0$ to infinity, it is unlikely that this infinite sum converges.  So in general, 1-loop contributions cannot be represented as
\be
\label{1loopsca}
V_1(\phi_{1c},\phi_{2c})=\fr{1}{2}\int\fr{d^4p}{(2\pi)^4}\log\left[
p^2+m^2(\phi_{1c},\phi_{2c})\right].
\end{equation}

The representation of one-loop contribution like the above result is only a stereotype application from the calculation results of the single-field case (i.e., from the result, $V_1(u_{c})=\fr{1}{2}\int\fr{d^4p}{(2\pi)^4}\log\left[		p^2+m^2(u_{c})\right]$ in Ref.~\cite{quiros}, $u_c$ is VEV of single scalar field). But the computation of the diagrams has been ignored. Although this is imprecise, if we unconditionally accept the one-loop contributions as Eq.~(\ref{1loopsca}), it is also a fairly general estimate of one-loop contributions when considering two fields at once. This is also a possible method today in the context of calculation Eq.~(\ref{dinh}). Nevertheless, it  is very difficult.

However, this representation is also true, if we interpret the above result as implying $\phi_{2c}=\kappa\phi_{1c}$, that is, the above result represents only one field $\phi_1$ or $\phi_2$; or ignoring all the diagrams where $\phi_1$ and $\phi_2$ are present at the same time. Therefore $V_1(\phi_{1c},\phi_{2c})\equiv V_1(\phi_{c})$ with $\phi_c^2=\phi^2_{1c}+\phi^2_{2c}$. This is a very good approximation that eliminates the difficulty of summing diagrams as shown in Fig.~\ref{fig5}. This approximation has been used in calculating decay channels or diagrams as in Refs.~\cite{tadashi,zhen, abdesslam}. Also, there are basis-independent methods for the two-Higgs-doublet model \cite{shinya,davidson}, it is possible to rewrite 2HDM model under one VEV. But in the sections of article, this approximation is only imprecise in the EWPT analysis.

The next interesting thing here is what if we could rewrite ${\cal L}\left(\phi_1(x),\phi_2(x)\right)={\cal L}\left(\phi_1(x)\right)+{\cal L}\left(\phi_2(x)\right)$, when expanding $\phi_i$ in terms of $\phi_{ic}$. At that point, the generating functional is rewritten as
\be
Z[J]=Z[J_1].Z[J_2],
\end{equation}
so that,
\be
W[J]=W[J_1]+W[J_2].
\end{equation}

In other words, the effective potential can be separated into two separate parts:
\begin{align}
\Ga[\overline{\phi}_1,\overline{\phi}_2]&=W[J]-\int d^4 x \fr{\de
W[J]}{\de J_i(x)} J_i(x)\\
&=W[J_1]-\int d^4 x \fr{\de
W[J_1]}{\de J_1(x)} J_1(x)+W[J_2]-\int d^4 x \fr{\de
W[J_2]}{\de J_2(x)} J_2(x)\\
&=\Ga[\overline{\phi}_1]+\Ga[\overline{\phi}_2]
\end{align}

This will make summing the diagrams easier, but they are not always separated like that. The cases of more than two fields are similarly constructed.
	
\end{document}